\documentclass[aps, prl, reprint, superscriptaddress, floatfix, amssymb]{revtex4-2}
\usepackage{xcolor,amsmath,url,graphicx, hyperref}
\usepackage{subcaption}

\usepackage[capitalize]{cleveref}
\crefname{equation}{Equation}{Equations}
\renewcommand{\figurename}{Fig.}

\makeatletter
\def\maketitle{
	\@author@finish
	\title@column\titleblock@produce
	\suppressfloats[t]}
\makeatother

\begin{document}
	\title{Recombination of localized quasiparticles in disordered superconductors}
	
	\author{Steven A. H. de Rooij}
	\email{s.a.h.de.rooij@sron.nl}
	\affiliation{SRON - Space Research Organisation Netherlands, Niels Bohrweg 4, 2333 CA Leiden, The
		Netherlands}
	\affiliation{Faculty of Electrical Engineering, Mathematics and Computer Science, Delft University of
		Technology, Mekelweg 4, 2628 CD Delft, the Netherlands}
		
	\author{Remko Fermin}
	\affiliation{Department of Materials Science and Metallurgy, University of Cambridge, 27 Charles Babbage Rd, Cambridge CB3 0FS, United Kingdom}
	\affiliation{Huygens-Kamerlingh Onnes Laboratory, Leiden University, PO Box 9504, 2300 RA Leiden, The Netherlands}
		
	\author{Kevin Kouwenhoven}
	\affiliation{SRON - Space Research Organisation Netherlands, Niels Bohrweg 4, 2333 CA Leiden, The
	Netherlands}
	\affiliation{Faculty of Electrical Engineering, Mathematics and Computer Science, Delft University of
	Technology, Mekelweg 4, 2628 CD Delft, the Netherlands}
	
	\author{Tonny Coppens}
	\affiliation{SRON - Space Research Organisation Netherlands, Niels Bohrweg 4, 2333 CA Leiden, The
	Netherlands}
	
	\author{Vignesh Murugesan}
	\affiliation{SRON - Space Research Organisation Netherlands, Niels Bohrweg 4, 2333 CA Leiden, The
		Netherlands}
	
	\author{David J. Thoen}
	\affiliation{SRON - Space Research Organisation Netherlands, Niels Bohrweg 4, 2333 CA Leiden, The
		Netherlands}
		
	\author{Jan Aarts}
	\affiliation{Huygens-Kamerlingh Onnes Laboratory, Leiden University, PO Box 9504, 2300 RA Leiden, The Netherlands}
		
	\author{Jochem J. A. Baselmans}
	\affiliation{SRON - Space Research Organisation Netherlands, Niels Bohrweg 4, 2333 CA Leiden, The
		Netherlands}
	\affiliation{Faculty of Electrical Engineering, Mathematics and Computer Science, Delft University of
		Technology, Mekelweg 4, 2628 CD Delft, the Netherlands}
	
	\author{Pieter J. de Visser}
	\affiliation{SRON - Space Research Organisation Netherlands, Niels Bohrweg 4, 2333 CA Leiden, The
		Netherlands}

\begin{abstract}
	Disordered superconductors offer new impedance regimes for quantum circuits, enable a pathway to protected qubits, and can improve superconducting detectors due to their high kinetic inductance and sheet resistance. The performance of these devices can be limited, however, by quasiparticles - the fundamental excitations of a superconductor. While experiments have shown that disorder affects the relaxation of quasiparticles drastically, the microscopic mechanisms are still not understood. We address this issue by measuring quasiparticle relaxation in a disordered $\beta$-Ta film, which we pattern as the inductor of a microwave resonator. We observe that quasiparticle recombination is governed by the phonon scattering time, which is faster than conventional recombination in ordered superconductors. We interpret the results as recombination of localized quasiparticles, induced by disorder, which first delocalize via phonon absorption. We analyze quasiparticle relaxation measurements on superconductors with different degrees of disorder and conclude that this phenomenon is inherent to disordered superconductors.
\end{abstract}
\maketitle

\begin{figure*}[t]
	\includegraphics[width=.9\textwidth]{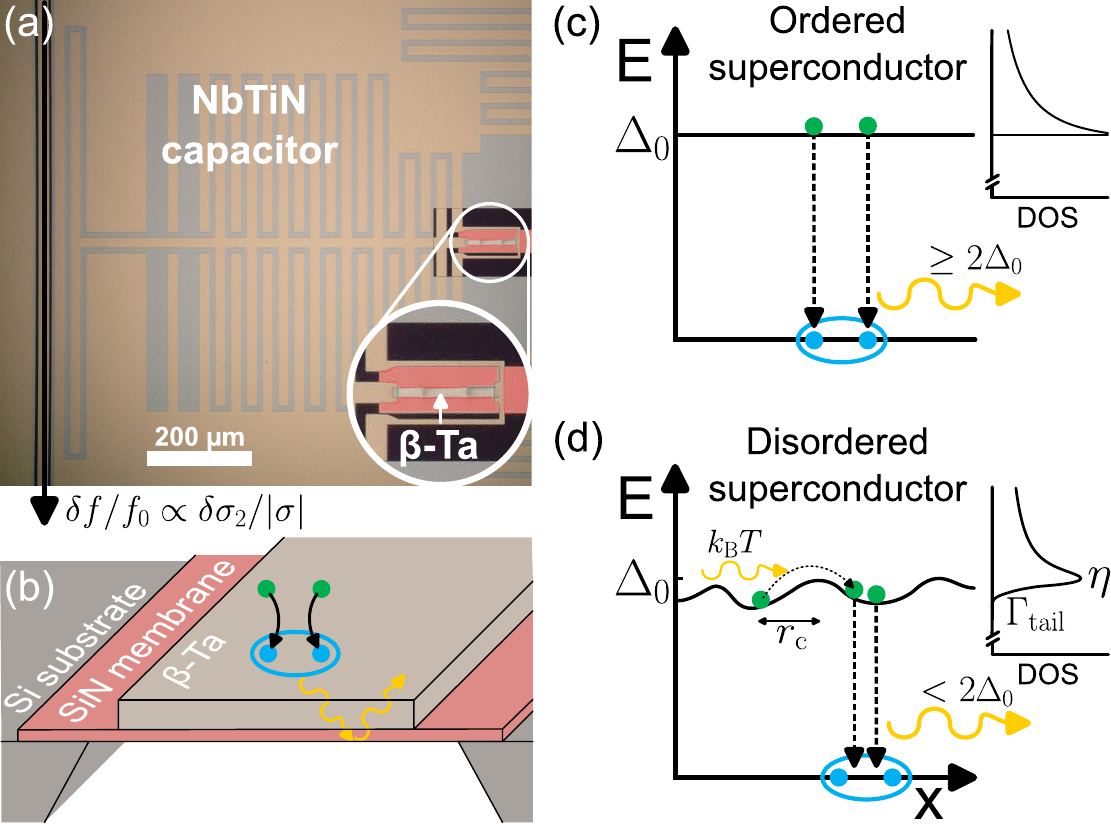}
	\caption{\label{fig:disrecsketch} \textbf{Quasiparticle fluctuation measurement and the inferred effect of disorder on quasiparticle recombination.} (a): Micrograph of the microwave resonator consisting of a NbTiN capacitor and $\beta$-Ta inductor on a membrane, which is highlighted in red. Inside the white circle the figure is two times enlarged. We measure fluctuations in resonance frequency ($\delta f$), which are proportional to complex conductivity fluctuations ($\delta\sigma_2$). (b): Sketch of the $\beta$-Ta inductor on the $110~\text{nm}$ thick SiN membrane. When two quasiparticles (green) recombine into a Cooper-pair (blue), a phonon is emitted (yellow curvy arrows). A change in the number of quasiparticles or Cooper-pairs changes $\sigma_2$, which we measure. The emitted phonon is trapped by the membrane, as indicated. We measure two resonators: one with the inductor on a SiN membrane (as sketched) and one with the SiN patch on solid Si substrate. (c): Sketch of traditional quasiparticle recombination with emission of a phonon with energy $\geq 2\Delta_0$, which can subsequently break a Cooper-pair. The BCS density of state (DOS) is sketched on the right, which has the same energy y-axis. (d): Sketch of quasiparticle recombination in disordered superconductors. Disorder can suppress the gap locally, inducing quasiparticle localization at a typical length scale of $r_\text{c}\sim\xi$ \cite{Bespalov2016a}. Quasiparticles can delocalize via absorption of a phonon, after which they relax and rapidly recombine on-site, emitting a phonon with less than $2\Delta_0$ energy. Since phonon absorption is slow in disordered superconductors \cite{Reizer1986,Pippard1955} and subsequent on-site recombination is fast \cite{Bespalov2016,Kozorezov2008}, delocalization limits this process. On the right, the position-averaged DOS is sketched with a broadened coherence peak, parameterized by $\eta$, and a sub-gap tail consisting of localized states, parameterized by $\Gamma_{tail}$ \cite{Feigelman2012,Larkin1972}.}
\end{figure*}
Disordered superconductors have a high resistance in their normal state, leading to a competition between electronic localization effects and the global phase coherence of superconductivity. A large amount of disorder can induce electronic granularity, pre-formed Cooper-pairs, ultimately causing a superconductor-to-insulator phase transition \cite{Sacepe2020, Bastiaans2021}. \\
Localization effects weaken the superfluid stiffness, thereby increasing the kinetic inductance ($L_\text{k}$) of the superconductor. This enables new regimes in high impedance quantum circuits and the development of protected qubits \cite{Grunhaupt2019}, as well as highly non-linear elements for parametric amplification \cite{Esposito2021a}. For superconducting single photon detectors \cite{Day2003, Goltsman2001}, the high normal state resistance of disordered superconductors increases the broadband photon absorption efficiency \cite{Kouwenhoven2022,Ma2024} and photon responsivity \cite{Leduc2010, Sidorova2023}.\\
Quasiparticles, the elementary excitations in a superconductor, play a central role in these applications. In quantum circuits, quasiparticles cause decoherence and microwave loss. Considerable efforts have been made to mitigate excess quasiparticles \cite{Martinis2021, Riwar2019, Karatsu2019,Baselmans2012a,Cardani2021} and to explain the non-vanishing quasiparticle density at low temperatures \cite{Bespalov2016a,deVisser2014b,Vepsalainen2020,Barends2011}. For superconducting radiation detectors, the relaxation of photon-generated quasiparticles dictates the detector performance \cite{Sergeev1996}. \\
As disorder increases, the limiting relaxation time at low temperatures becomes shorter \cite{Barends2009a}. In contrast, quasiparticle relaxation times on the order of seconds have been measured in granular Al, which is highly disordered due to oxidized grain boundaries \cite{Grunhaupt2018}. Furthermore, when exciting disordered superconductors with electromagnetic radiation, an anomalous quasiparticle response has been measured \cite{Gao2012,Bueno2014} and microwave loss typically increases with $L_\text{k}$ \cite{Moshe2020}. This hinders the use of disordered superconductors in quantum circuits and superconducting radiation detectors. These observations suggest that disorder affects quasiparticle relaxation, but the underlying mechanisms remain poorly understood.\\
In this work, we demonstrate that localization effects govern the relaxation of quasiparticles in disordered superconductors. Disorder induces local gap inhomogeneities that serve as localization sites where quasiparticles rapidly recombine. Consequently, the relaxation process is not governed by recombination of mobile quasiparticles, as in ordered superconductors (\cref{fig:disrecsketch}(c)), but rather by delocalization and subsequent on-site recombination with a localized quasiparticle (\cref{fig:disrecsketch}(d)). This leads to a shorter relaxation time with a weaker temperature dependence than in the absence of localization effects. Such a short relaxation time will help to mitigate excess quasiparticles in quantum circuits \cite{Martinis2021}. On the other hand, at low temperatures and quasiparticle generation rates, a background of localized quasiparticles may induce additional microwave loss \cite{Grunhaupt2018} and limit the response of superconducting detectors \cite{Bueno2014,Gao2012}.
\section*{Results}
\begin{figure*}[t]
	\includegraphics[width=\textwidth]{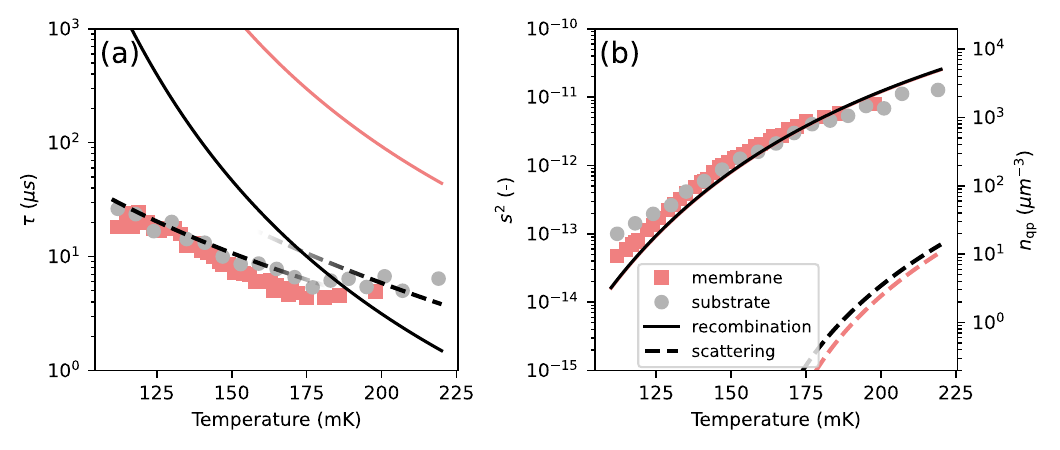}
	\caption{\label{fig:results} \textbf{Measured relaxation time and variance compared with quasiparticle fluctuation theory.} The gray circles and red squares are determined from a fit to the measured fluctuation spectra (\cref{fig:specs}), which gives the relaxation time (a) and variance (b) of the quasiparticle fluctuation. The red squares are for the inductor on a SiN membrane (see \cref{fig:disrecsketch}(b)) and the gray circles on a SiN/Si solid substrate, as indicated by the legend. Error bars from statistical fit errors are smaller than the data points. The dashed lines in (a) are calculations for electron-phonon scattering time, \cref{eq:tauscat}, which is 2D for $T<165~\text{mK}$ \cite{Devereaux1991a} and 3D for higher temperatures \cite{Reizer1986}. The dashed lines in (b) are the variance of quasiparticle fluctuations for scattering events. The solid lines are for recombination. The red solid line in (a) includes the expected increase in relaxation time due to phonon trapping by the membrane (\cref{fig:disrecsketch}(b)). These curves are calculated with the parameters in \cref{tab:params} without fit parameters. The right axis in (b) gives the quasiparticle density corresponding to the variance of recombination events (the solid line).}
\end{figure*}
\subsection{Fluctuation measurements}
We measure the quasiparticle relaxation time ($\tau$) and density ($n_{\text{qp}}$) in a $40~\text{nm}$ thick disordered $\beta$-Ta film \cite{Read1965}, which is patterned as the inductor in a microwave resonator. The device is shown in \cref{fig:disrecsketch}(a) and the film parameters are given in \cref{tab:params}. This $\beta$-Ta film is disordered as the Ioffe-Regel parameter, $k_\text{F}l$, is close to one, with $k_\text{F}$ the Fermi wave number and $l$ the electronic mean free path. We therefore expect localization effects to become important. Moreover, $ql$ is much smaller than unity, with $q$ the phonon wave number of a phonon with energy $2\Delta_0$, which is the BCS superconducting gap energy at low temperatures. This means the film is disordered with respect to electron-phonon interactions \cite{Reizer1986}. The characterization of the film is outlined in the Methods section.\\
We cool the resonators to bath temperatures ranging from $20~\text{mK}$ to $300~\text{mK}$. For each bath temperature, we drive the resonator at its resonance frequency and measure time streams of the complex microwave transmission in a homodyne setup (see Methods). From that, we extract the Power Spectral Density (PSD) of the fluctuations in $\sigma_2$, $S_{\delta\sigma_2/|\sigma|}(\omega)$, where $\sigma=\sigma_1 - i \sigma_2$ is the complex conductivity. This is equivalent to measuring the resonator frequency noise. These fluctuations are characterized by a Lorentzian spectrum \cite{Wilson2004}, as is visible in the measured PSDs in \cref{fig:specs}. We fit a Lorentzian spectrum and extract the variance, $s^2$, and relaxation time, $\tau$, as described in the Methods section.\\
The measured $\tau$ and $s^2$ are shown in \cref{fig:results}(a) and (b). For higher and lower bath temperatures than shown in \cref{fig:results}, the quasiparticle fluctuations were obscured by other noise sources such as $1/f$ and amplifier noise.

\subsection{Phonon scattering limited quasiparticle relaxation}
We compare these measured values to the theory for quasiparticle relaxation. At low temperatures, $k_\text{B}T\ll\Delta_0$ with $T$ the bath temperature and $k_\text{B}$ the Boltzmann constant, $\tau$ is governed by electron-phonon interactions, because $n_{\text{qp}}$ is low compared to the normal state carrier density \cite{Kaplan1976}. As a check, we compared the electron-electron and electron-phonon interaction times \cite{Reizer2000} and find that the electron-phonon interaction has the shortest relevant timescales in our system, as outlined in the Supplementary Section 1.\\ 
There are two inelastic electron-phonon processes that contribute to the relaxation of quasiparticles: recombination and scattering. 

During a recombination event, two quasiparticles recombine into a Cooper-pair and emit a phonon with energy $\Omega\geq2\Delta_0$, as illustrated in \cref{fig:disrecsketch}(c). Quasiparticle generation is the same process, but time-reversed. The recombination time, $\tau_{\text{rec}}$, is proportional to $1/n_{\text{qp}}$, which exponentially deceases with temperature, i.e. \cite{Kaplan1976, Reizer1986} 
\begin{equation}\label{eq:taurec}
	\tau_{\text{rec}}(T)=\tau_0^{\text{rec}} \sqrt{\frac{T_\text{c}}{T}} e^{\Delta_0/k_\text{B}T}. 
\end{equation}
$T_\text{c}$ is the critical temperature and $\tau_0^{\text{rec}}$ is a proportionality constant that does not depend on temperature, but does depend on disorder and dimensionality (see Methods).\\
In our experiment, we probe the entire inductor volume and therefore measure the relaxation of an ensemble of quasiparticles. The bulk recombination time is given by \cite{Rothwarf1967},
\begin{equation}\label{eq:taurecstar}
	\tau_{\text{rec}}^*=\tau_{\text{rec}}(1+\tau_{\text{esc}}/\tau_{\text{pb}})/2,
\end{equation}
where $\tau_{\text{esc}}$ is the escape time and $\tau_{\text{pb}}$ is the pair-breaking time of a $2\Delta_0$ phonon. The last division by two in \cref{eq:taurecstar} describes the pair-wise nature of recombination. The factor in parenthesis is the phonon trapping factor and takes into account that the emitted $2\Delta_0$-phonon can subsequently break another pair. Quasiparticles in an ordered superconductor, like Al and $\alpha$-Ta with $ql\gg1$, relax with a relaxation time given by \cref{eq:taurecstar}, which increases when phonon trapping is enhanced \cite{Kaplan1976, deRooij2021, Barends2009a}.\\
For the substrate case, with $\beta$-Ta on a SiN/Si substrate, we estimate the phonon trapping factor to be $2.0$ from the acoustic mismatch theory set out in Ref. \cite{Kaplan1979,Eisenmenger1976}. Using the parameters of the film (\cref{tab:params}), which are independently measured without fitting to the data, we obtain the solid black line in \cref{fig:results}(a). Details on the calculation of the curves in \cref{fig:results} are presented in Supplementary Section 1.\\
We increase $\tau_{\text{esc}}$ experimentally by etching the Si under the SiN patch of the resonators, such that the $\beta$-Ta inductor is suspended on a $110~\text{nm}$ thick membrane. This shown in \cref{fig:disrecsketch}(b). Via a simple geometric calculation (Supplementary Section 1), we estimate that the membrane results in an increase of $\tau_{\text{esc}}$ by factor $58$ compared to the substrate case. That would increase $\tau_{\text{rec}}^*$ by a factor $29$, which is shown in \cref{fig:results}(a) as the red solid line. We observe that the quasiparticle relaxation time is identical for both the substrate and membrane case, which is not in agreement with conventional quasiparticle recombination. Moreover, the temperature dependence of the relaxation time is much weaker than the conventional exponential temperature dependence (\cref{eq:taurecstar}).

The dashed lines in \cref{fig:results}(a) are for quasiparticle-phonon scattering. During a scattering event, a quasiparticle absorbs a thermal phonon of energy $\sim k_\text{B}T$. The inelastic time related to scattering, $\tau_{\text{scat}}$, is proportional to the phonon occupation and therefore follows a power law versus temperature, 
\begin{equation}\label{eq:tauscat}
	\tau_{\text{scat}}(T) =\tau_0^{\text{scat}} \left(\frac{T}{T_\text{c}}\right)^{-n},
\end{equation}
with $2 \leq n \leq 9/2$. $n$ and $\tau_0^{\text{scat}}$ depend on disorder and the electron and phonon dimensionality \cite{Devereaux1991a, Reizer1986, Kaplan1976, Sergeev2000}. In our film, thermal phonons are 2D for $T\lesssim165~\text{mK}$ and 3D for higher temperatures. This changes the exponent $n$ from $9/2$ in 3D \cite{Reizer1986} to $7/2$ in 2D \cite{Devereaux1991a}. Using the parameters from \cref{tab:params} results in the dashed lines in \cref{fig:results}(a). We divide \cref{eq:tauscat} by a factor 2 in \cref{fig:results}(a) to account for pair-wise recombination after the phonon scattering event, similar to \cref{eq:taurecstar}.\\
The agreement between the measured quasiparticle relaxation time and the power-law temperature dependence of phonon scattering, indicates that quasiparticle relaxation is governed by phonon scattering. This is the main experimental result of this work and is in sharp contrast to ordered superconductors, such as Al and $\alpha$-Ta \cite{Barends2009a,deRooij2021}.

\subsection{Verification of quasiparticle generation-recombination fluctuations}
To examine which relaxation process we observe, we compare the measured variance to the calculated variance of quasiparticle generation-recombination and scattering fluctuations in $\sigma_2$. This is shown in \cref{fig:results}(b). We calculate the generation-recombination noise variance with $s^2=(n_{\text{qp}}/V) (d(\sigma_2/|\sigma|)/dn_{\text{qp}})^2$ \cite{Wilson2004}, where $V$ is the $\beta$-Ta volume and the factor in parentheses is the responsivity of the complex conductivity to changes in the quasiparticle density. This results in the solid line in \cref{fig:results}(b). This variance is equal for the membrane and substrate case, because it does not depend on phonon trapping. The responsivity factor is approximately constant with temperature ($<2~\%$ deviation in our measurement regime) and known from theory \cite{Mattis1958,Gao2008}, which allows us to directly measure the quasiparticle density from the variance \cite{deVisser2011}. The right axis of \cref{fig:results}(d) gives the quasiparticle density corresponding to the variance.\\
The dashed lines in \cref{fig:results}(b) give the variance for scattering interactions. It is orders of magnitude lower than the variance for recombination events, because the energy difference of a scattering event is much less than that of a recombination event, $k_\text{B}T\ll\Delta_0$. The variance of the fluctuations clearly corresponds to the recombination of thermal quasiparticles \cite{Wilson2004}. We attribute the small deviation from the thermal line at low temperatures to microwave read-out power effects \cite{deVisser2014b}, see also \cref{fig:Pread}. \\
We performed the same analysis to the measured dissipative fluctuations, $\delta\sigma_1$, which are shown in \cref{fig:s1results}. The results show the same phenomenology as \cref{fig:results}: a relaxation time governed by phonon scattering and a generation-recombination noise variance. This shows that we measure quasiparticle recombination events \cite{Wilson2004,deVisser2011}. \\
To summarize the above, we observe that the quasiparticle relaxation in this disordered film is governed by fast quasiparticle recombination with a time scale given by phonon scattering, which is not affected by phonon trapping. This is in sharp contrast to ordered superconductors \cite{deRooij2021}.
\subsection{Recombination of localized quasiparticles}
We explain these observations by localized quasiparticles that delocalize via phonon absorption. A delocalized quasiparticle relaxes again to another localization site and subsequently recombines on-site (\cref{fig:disrecsketch}(d)). Phonon absorption is relatively slow in a disordered ($ql\ll 1$) metal \cite{Pippard1955} and the subsequent on-site recombination is fast, since the quasiparticles relax to the same location \cite{Bespalov2016a}. Therefore, phonon absorption limits the relaxation time, as observed in \cref{fig:results}(a). Even in the membrane case, where recombination should be slow due to phonon trapping, phonon absorption limits the relaxation time. The reason for this could be the following. The phonon that is emitted during on-site recombination has an energy below $2\Delta_0$, since the delocalized quasiparticle relaxes further into a localization site (\cref{fig:disrecsketch}(d)). The chance for such a phonon to break a Cooper-pair is very low, because the density of localized states with $E<\Delta_0$ is very low (inset of \cref{fig:disrecsketch}(d)). Therefore, the phonon trapping effect for on-site recombination should be largely reduced \cite{Kozorezov2008}. This is consistent with \cref{fig:results}, where we observe the relaxation time to be equal to the scattering time for both the membrane and substrate case.\\
To support this interpretation of the results, we modeled the quasiparticle fluctuations with the master equation approach described in Ref. \cite{Wilson2004}. We describe the quasiparticles as localized such that their recombination rate depends on the average distance between two quasiparticles, as set out in Ref. \cite{Bespalov2016a}. For vanishing temperatures, this results in an exponentially long relaxation time and an excess quasiparticle density \cite{Bespalov2016a}. We however measure at a finite temperature where quasiparticles can delocalize via inelastic phonon scattering. We therefore include a second quasiparticle level to describe mobile quasiparticles, and set the delocalization time to $\tau_{\text{scat}}$ from \cref{eq:tauscat}. We assume the localization time to be much shorter, such that the mobile quasiparticle density is small. We include an on-site recombination term, where a mobile quasiparticle relaxes and recombines on-site with a localized quasiparticle (\cref{fig:disrecsketch}(d)). We set the characteristic time for on-site recombination, $\tau_0^{\text{os}}$ c.f. $\tau_0^{\text{rec}}$ in \cref{eq:taurec}, such that on-site recombination is much faster than localization.\\
The resulting fluctuation spectra for the total number of quasiparticles are shown in \cref{fig:specs}(c) and (d). These calculated spectra follow the observed behavior. When analytically examining this model, we see that in the limit of fast on-site recombination, the quasiparticle relaxation time is given by $\tau_{\text{scat}}/2$ (the dashed lines in \cref{fig:results}(a)) and the variance is equal to $n_{\text{qp}}$ (black line in \cref{fig:results}(b)). For details on the model calculations, see Supplementary Section 5.

The proposed origin of the localized states is local gap variations induced by disorder. This is described by the theory of Refs. \cite{Feigelman2012, Skvortsov2013}, which extends the Larkin-Ovchinnikov inhomogeneous pairing theory \cite{Larkin1972}. Such gap variations have been measured in highly disordered TiN \cite{Sacepe2008}. The effect of inhomogeneous pairing on the density of states is two-fold: (1) the coherence peak is broadened with a pair-breaking parameter $\eta$, which is equivalent to the Abrikosov-Gor'kov description of magnetic impurities \cite{Abrikosov1960}, and (2) the density of states acquires an exponential subgap tail of localized states, characterized by $\Gamma_{\text{tail}}$. This is sketched in the inset of \cref{fig:disrecsketch}(d). The values of $\eta$ and $\Gamma_{\text{tail}}$ depend on the strength of the local gap variations.\\
Because the thickness of our $\beta$-Ta film is on the order of the coherence length, $\xi$ (see Methods), we take both Coulomb induced mesoscopic fluctuations and finite thickness effects into account and obtain $\eta\approx 2.4\times10^{-5}$ \cite{Feigelman2012}. To estimate $\Gamma_{\text{tail}}$, we consider our film quasi-2D, in which case Coulomb-enhanced mesoscopic fluctuations dominate and give $\Gamma_{\text{tail}}/\Delta_0\approx 1.5\times10^{-4}$ \cite{Feigelman2012}. Quasiparticle states in this subgap tail overlap and quasiparticles in these states will relax further. The effective localization radius at which this process stops can be estimated from $\Gamma_{\text{tail}}$ and $\eta$ and is on the order of $\xi$ \cite{Bespalov2016a}. With the values mentioned above, $r_\text{c}\approx 3.5 \xi \approx 65~\text{nm}$. This gives an estimate of density of localized quasiparticle states of $\tilde{n}_\text{qp}^\text{loc}=3/(4\pi r_\text{c}^3)\approx8.5\times10^{2}~\mu\text{m}^{-3}$. For details see Supplementary Section 6.\\
Comparing this number to the right axis of \cref{fig:results}(b) we conclude that the thermally generated quasiparticles localize due to gap variations for temperatures $T<178~\text{mK}$. This leads to the situation in \cref{fig:disrecsketch}(d). For higher temperatures, the thermal quasiparticle density is larger than the density of localized states. This alters the quasiparticle localization dynamics and might be related to the plateau we observe in \cref{fig:results}(a) for $T>175~\text{mK}$.
\section{Discussion}
\begin{figure*}[t]
	\includegraphics[width=.8\textwidth]{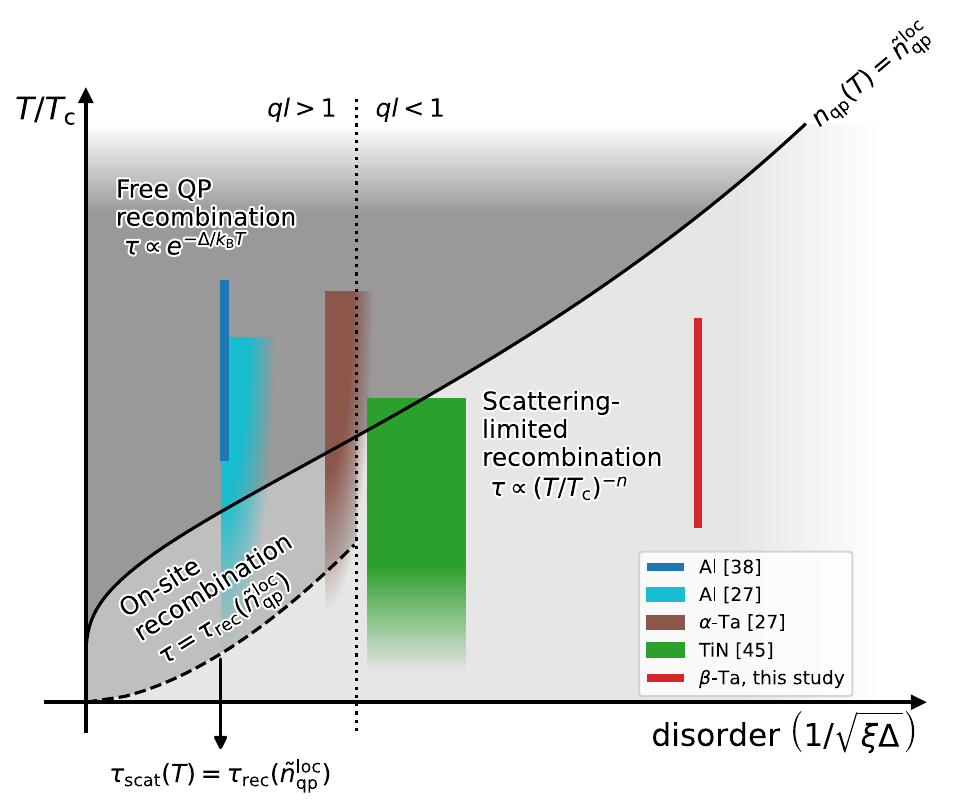}
	\caption{\label{fig:disorder diagram} \textbf{Different regimes of quasiparticle recombination as a function of temperature and disorder.} Each regime (indicated in a shade of gray) is characterized by a different temperature dependence of the quasiparticle (QP) relaxation time. The x-axis is chosen such that the dotted vertical line, $ql=1$ for $2\Delta_0$ phonons, is approximately on the same point for different superconductors. For simplicity, we take $ql$ for $2\Delta_0$ phonons only in this sketch, such that on the right side of the $ql=1$ dotted line both recombination and scattering interactions are disordered. The solid line gives the values at which the thermal quasiparticle number equals the number of localized quasiparticle states, which we estimate as $\tilde{n}_{\text{qp}}^{\text{loc}}=3/(4\pi\xi^3)$. Below this solid line, quasiparticles will localize and recombine on-site. Below the dashed line the scattering time, \cref{eq:tauscat}, is longer than the recombination time, \cref{eq:taurec}, at a quasiparticle density equal to $\tilde{n}_{\text{qp}}^{\text{loc}}$, i.e. $\tau_{\text{rec}}(\tilde{n}_{\text{qp}}^{\text{loc}})$ \cite{Kaplan1976,Reizer1986}. In that regime, delocalization via phonon absorption limits quasiparticle recombination (\cref{fig:disrecsketch}(d)), which is indicated in light gray. For $ql<1$ (the right side of the dotted line), phonon absorption is always slower than recombination at $\tilde{n}_{\text{qp}}^{\text{loc}}$. The colored areas give the experimental conditions of Refs. \cite{deRooij2021,Barends2009a,Coumou2013a} and \cref{fig:results}. The fade at lower temperatures for Refs. \cite{Barends2009a,Coumou2013a} are added because these are measurements via a pulsed excitation, which does not give information on the quasiparticle temperature. In Ref. \cite{Barends2009a}, disorder is increased by implanting ions. There is however no data on how much $1/\sqrt{\xi\Delta}$ increased in these films, which we indicated by an additional fade to higher disorder. The recombination regimes are consistent with the observed temperature dependence of the relaxation time in the respective experiments.}
\end{figure*}
Following the same line of reasoning, we expect to see the same phenomenology in other disordered superconductors. Indeed, a weak temperature dependence of the quasiparticle relaxation time is also observed in TiN for different levels of disorder \cite{Coumou2013a}, which was not understood before. The electron-phonon scattering time at $T_\text{c}$ has also been measured for these films \cite{Kardakova2015}, which provides an accurate estimate of $\tau_0^{\text{scat}}$ and $\tau_0^{\text{rec}}$ in \cref{eq:taurec,eq:tauscat}. We analyzed those results and find that static scatterers cause a power law temperature dependence of the phonon scattering time with $n=2$ \cite{Sergeev2000}. Taking this into account, the relaxation times for these TiN films follow both the temperature and disorder dependence of electron-phonon scattering, similar to \cref{fig:results}(a), see Supplementary Section 3. With $r_\text{c}\sim\xi^{\text{TiN}}$ \cite{Bespalov2016a} these films are in the same regime as $\beta$-Ta, where $n_{\text{qp}}(T)<\tilde{n}_{\text{qp}}^{\text{loc}}$ and $ql<1$. Therefore, also in these films thermal quasiparticles localize and the recombination process is limited to delocalization via phonon absorption (\cref{fig:disrecsketch}(d)).\\
Even in an ordered superconductor a small amount of disorder could induce quasiparticle localization at low temperatures \cite{Bespalov2016a}. In Ref. \cite{deRooij2021}, we observed for Al that there is no effect of phonon trapping on the relaxation time in the saturation regime at low temperatures. We explained that by the presence of an excess number of localized quasiparticles. We estimate that the density of localized quasiparticle states for the quasi-2D Al film is given by, $\tilde{n}_{\text{qp}}^{\text{loc}}\simeq 1/(\pi r_\text{c}^2 d)\approx 160~\mu \text{m}^{-3}$, where we take $r_\text{c}\sim \xi^{\text{Al}}\approx 200~\text{nm}$. If all these states are filled due to a non-equilibrium generation process, the quasiparticle density corresponds to the observed saturation time of $1~\text{ms}$ in Ref. \cite{deRooij2021}. An increase in disorder increases the number of localized quasiparticle states and decreases this saturation time. This is experimentally shown in \cite{Barends2009a} for Al and $\alpha$-Ta. Furthermore, the most disordered Al film in Ref. \cite{Barends2009a} shows a weaker temperature dependence of $\tau$, which points towards the same phenomenology as observed in \cref{fig:results}(a).\\
In \cref{fig:disorder diagram}, we sketch these different quasiparticle recombination regimes. Above the solid line the thermal quasiparticle density is larger than the density of localized quasiparticle states. In that regime, the relaxation time follows the free quasiparticle recombination time, $\tau_{\text{rec}}^*(T)$ given by \cref{eq:taurecstar,eq:taurec} and shown in \cref{fig:disrecsketch}(c). Below the solid line, quasiparticles recombine within a localization site. The relaxation time saturates to the recombination time at a quasiparticle density equal to the number of localization sites, $\tilde{n}_{\text{qp}}^{\text{loc}}$, as observed in Refs. \cite{deRooij2021,Barends2009a}. However, if the phonon scattering time is longer than this saturation time, quasiparticle localization via phonon absorption limits the relaxation and the temperature dependence of the relaxation time is altered to the power-law of $\tau_{\text{scat}}$, \cref{eq:tauscat}. This regime is indicated as the light gray area in \cref{fig:disorder diagram}. The results of Refs. \cite{deRooij2021,Barends2009a,Coumou2013a}, which were previously unexplained, and \cref{fig:results} are shown in the appropriate disorder and temperature regimes and are consistent with the observed temperature dependence of the measured relaxation times.

A saturation of the quasiparticle relaxation time requires a non-equilibrium quasiparticle generation process such as microwave readout power \cite{deVisser2014b}, cosmic rays \cite{Karatsu2019}, radioactivity \cite{Cardani2021} and stray light \cite{Baselmans2012a}. In \cref{fig:disorder diagram}, this can be viewed as a saturation of $T/T_\text{c}$, when $T$ is an effective quasiparticle temperature. We assume that there is a number of excess quasiparticles present in our experiment due to the continuous microwave read-out \cite{deVisser2014b}, which is small compared to the thermal quasiparticle density. To verify this, we show the results for $\tau$ and $s^2$ obtained in the same way, but for higher read-out powers in \cref{fig:Pread}. We observe that the excess quasiparticle generated by read-out power is minimized at a read-out power equivalent to a number of photons in the resonator of $\bar{n}_{\text{ph}}\approx6\times10^4$, which was used to obtain \cref{fig:results} (see Supplementary Section 7 for an estimation of the read-out power effects at this read-out power). If we would reduce the bath temperature and minimize this generation process further, quasiparticles can become fully localized, which enhances the relaxation time exponentially \cite{Bespalov2016a}. For example, a relaxation time in granular Al on the order of seconds has been measured with $\bar{n}_{\text{ph}}\approx 1 - 300$ at $25~\text{mK}$ \cite{Grunhaupt2018}. This relaxation time decreases with increasing $\bar{n}_{\text{ph}}$, indicating a saturated excess quasiparticle density that is sustained by microwave power. To verify that this originates from the localized quasiparticle dynamics presented in Ref. \cite{Bespalov2016a}, an experiment that independently probes both the relaxation rate and the quasiparticle effective temperature is needed. Such an experiment would be similar to this work, but at much lower temperatures and generation rates.\\
During the review process of this Article some related studies became available, which we reference here for completeness \cite{Fischer2025,Gurra2025,Charpentier2025a}. 

To conclude, we showed that quasiparticle relaxation in a disordered superconductor (with $ql<1$) is governed by quasiparticle localization. After delocalization via phonon absorption, quasiparticles relax and recombine rapidly on-site (\cref{fig:disrecsketch}(d)), which enhances the relaxation rate. This impacts the performance of superconducting devices. For radiation detectors that measure the presence of photon-generated quasiparticles, the signal will be reduced as a result of the faster relaxation time. This presents a trade-off between photon absorption efficiency (increasing with disorder) and signal (decreasing with disorder). In quantum circuits, the short relaxation time will help eliminate excess quasiparticles at critical elements such as Josephson junctions, thereby improving coherence \cite{Martinis2009}. At low temperatures and small quasiparticle generation rates, localized quasiparticles could however increase microwave loss \cite{Larkin1972, Grunhaupt2018}. Therefore, the fundamentally different quasiparticle dynamics presented in this work must be considered when implementing disordered superconductors in quantum circuits and radiation detectors.

\section{Methods}
\begin{table*}[t]
	\caption{\label{tab:params} \textbf{Geometry, electronic properties and phonon properties for $\beta$-Ta.} The electronic properties are obtained via a resistance versus temperature, a Hall resistance versus magnetic field and an upper critical field versus temperature measurement, as set out in Supplementary Section 2.}
	\begin{ruledtabular}
		\begin{tabular}{ccc|cccccc|cccc}
			\multicolumn{3}{c}{Geometry} & \multicolumn{6}{c}{Measured electronic properties} & \multicolumn{4}{c}{Values from Ref. \cite{Abadias2019}}\\ \hline
			$d$ & $W$   & $L$ & $T_\text{c}$ & $\rho_\text{N}(T=1~\text{K})$ & $D$ & $n_\text{e}$ &  &  & $\hat{\rho}$  & $c_\text{L}$ & $c_\text{T}$ & $T_\text{D}$\\ 
			(nm) & ($\mu$m) & ($\mu$m) & (K) &  ($\mu\Omega$cm) & ($\text{cm}^2/\text{s}$) &  ($\mu$m\textsuperscript{-3}) & RRR  & $k_\text{F} l$ &  ($g/cm\textsuperscript{3}$) & (km/s) & (km/s) & (K) \\
			40  & 10  & 90  & 0.87 & 206 & 0.74 & $9.3\times10^{10}$ & 1.03 & 4.2 & 16.6  & 4.34 &  1.73  & 221 \\
			
		\end{tabular}
	\end{ruledtabular}
\end{table*}
\subsection{Device design and fabrication}
The capacitive part of the resonator is an interdigitated capacitor (IDC), with $20~\mu\text{m}$ wide fingers and $10~\mu\text{m}$ wide gap. It is patterned using a SF\textsubscript{6} Reactive Ion Etch (RIE) in a 150 nm thick NbTiN film \cite{Thoen2017} with a critical temperature of 14.0 K and resistivity of $260~\mu\Omega\text{cm}$. This design minimizes noise from Two Level Systems (TLS) \cite{Gao2008c} and ensures that resonance frequency is highly sensitive to changes in the inductive $\beta$-Ta section, where the current density is high. The inductor is a strip of $\beta$-Ta, see \cref{tab:params}, patterned with a SF\textsubscript{6} RIE etch. We place the $\beta$-Ta film on a SiN patch, which serves as a membrane for one of the resonators after the Si wafer is etched away from the backside using KOH, see \cref{fig:results}(a) and (b). The inductor is shorted at the end to the NbTiN ground plane to make a quarter wave resonator. Quasiparticles in the $\beta$-Ta are confined to the inductor volume due to the higher superconducting gap of the NbTiN.\\
The resonance frequency is 5.1 GHz for the membrane resonator and 5.3 GHz for the substrate resonator, which is set by the finger lengths of the IDC. We set the coupling quality factor, $Q_\text{c}$, to ${\sim}10,000$ by tuning the length of the coupling bar next to the read-out line. The internal quality factor, $Q_\text{i}$, at $20~\text{mK}$ is approximately $400,000$. 

\subsection{Film characterization}
To obtain a measure of disorder and electron and phonon dimensionalities, we performed a measurement of the resistance versus temperature, which provides the normal state resistivity, $\rho_\text{N}$, and $T_\text{c}$, the Hall resistance at $1~\text{K}$, which provides the charge carrier density, $n_\text{e}$, and the upper critical field as a function of temperature, which provides the diffusion constant, $D$. The results can be found in \cref{tab:params} and the details of these measurements are given in Supplementary Section 2. The Ioffe-Regel parameter, $k_\text{F} l$, with $k_\text{F}$ the Fermi wavenumber and $l$ the electron mean free path, is of order unity. Therefore, the film is electronically disordered and we expect localization effects become important. Refs. \cite{Feigelman2012,Skvortsov2013} describe what electronic disorder does to the superconducting state: it introduces a broadening of the density of states and a subgap tail, consisting of localized quasiparticle states (see \cref{fig:disrecsketch}(d)).\\
For phonon-mediated superconductivity, disorder is characterized with respect to the electron-phonon interaction, i.e. $ql$, with $q$ the phonon wave number. We calculate $ql$ at two phonon energies: $2\Delta_0$, corresponding to recombination phonons and $k_\text{B}T$, typical phonon energies for scattering. We use the mass density, $\hat{\rho}$, and the longitudinal and transverse phonon velocities, $c_\text{L}$ and $c_\text{T}$, from Ref. \cite{Abadias2019}, see \cref{tab:params}. We only consider transverse phonons because these are the fastest the relaxation rates in our case, as $(c_\text{T}/c_\text{L})^3\ll 1$ \cite{Sergeev2000,Reizer1986}. With that, we calculate $q(2\Delta_0)l=0.070$ and $q(0.2~\text{K})l=0.0046$: both much smaller than 1 so the film is disordered with respect to electron-phonon interactions. We therefore use quasiparticle relaxation time calculations of Reizer and Sergeyev \cite{Reizer1986} and Devereaux and Belitz \cite{Devereaux1991a} which are in the disordered limit ($ql\ll 1$), instead of the widely used pure limit ($ql\gg 1$) results of Kaplan et al. \cite{Kaplan1976}. The main difference is that the scattering time (\cref{eq:tauscat}) has a steeper temperature dependence in the disordered case ($n\rightarrow n + 1$) and the proportionality constants in \cref{eq:taurec,eq:tauscat}, $\tau_{\text{rec}}^0$ and $\tau_{\text{scat}}^0$, become proportional to $1/\rho_\text{N}$, which reflects the weakening of the electron-phonon coupling as disorder increases \cite{Pippard1955}.

For electronic dimensionality, the dirty limit coherence length divided by the film thickness is, $\xi/d=\sqrt{l\xi_0}/d=0.39< 1$, with $l=0.30~\text{nm}$ and $\xi_0=0.79~\mu\text{m}$, the Pippard coherence length. Therefore, the $\beta$-Ta film is a 3D superconductor, although it is close to 2D. \\
For the phonon dimensionality, we compare the phonon wavenumbers to the film thickness, $qd$. For recombination phonons, the film is 3D ($q(2\Delta_0)d=9.3>1$) and we can use the results of Ref. \cite{Reizer1986} for $\tau_{\text{rec}}^0$. For scattering phonons, the film is 2D for $T<165~\text{mK}$ (when $q(k_\text{B}T)d\leq1/2$) and 3D for higher temperatures. The phonon dimensionality dictates the temperature dependence, while the electronic dimensionality dictates the disorder dependence \cite{Devereaux1991a}. We therefore use the 2D results of Ref. \cite{Devereaux1991a} with $n=7/2$ for the low temperature regime and the 3D result of Ref. \cite{Reizer1986} with $n=9/2$ for the high temperatures.

\subsection{Setup}
The sample is cooled in a dilution refrigerator, shielded from stray light with a box-in-a-box setup \cite{Baselmans2012a}. Magnetic interference is reduced by a factor $10^{-6}$ by a CRYOPHY and superconducting niobium shield. The forward transmission measurement is performed in a homodyne setup. The microwave signal is attenuated at each temperature stage before it reaches the sample. After the signal passed the sample, it is amplified by a HEMT amplifier at 3 K and by a room temperature amplifier before it is mixed with the original microwave signal by an IQ mixer. For details on all the components see Ref. \cite{deVisser2014}. 

\subsection{Fluctuation measurement}
\begin{figure*}[t]
	\includegraphics[width=\textwidth]{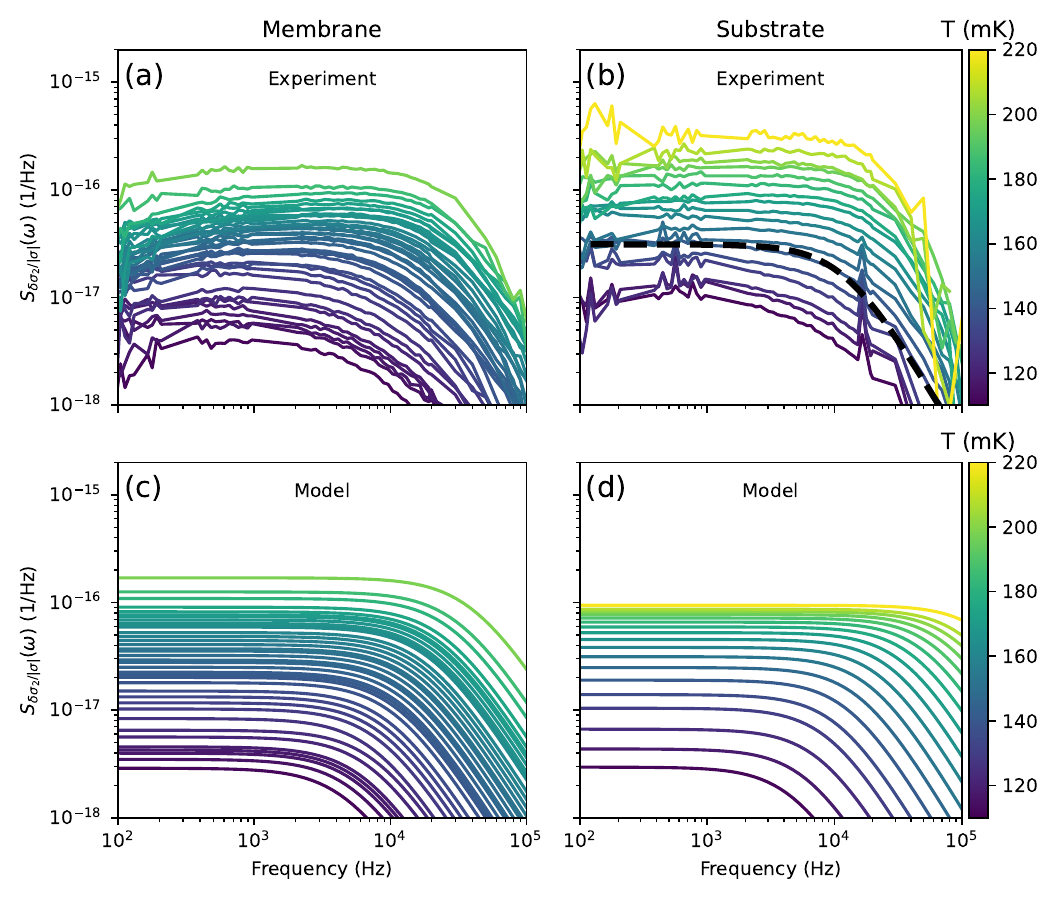}
	\caption{\label{fig:specs} \textbf{Measured and simulated power spectral densities of the complex conductivity.} Different colors are for different bath temperatures, as indicated by the colorbar. Other noise sources, such as amplifier noise and $1/f$-noise have been subtracted from the measured spectra in (a) and (b), as explained in the Supplementary Section 4. The dashed black line in (b) is an example fit using \cref{eq:Sfit}. (c) and (d) are calculated power spectral densities for quasiparticle fluctuations at the same bath temperatures as the measurements. The quasiparticle fluctuation spectra are calculated with the framework described in Ref. \cite{Wilson2004}, including the localized quasiparticle dynamics described in Ref. \cite{Bespalov2016a}. The quasiparticle density fluctuations are analytically converted to $\sigma_2/|\sigma|$-fluctuations. For details on the model and calculations, see Supplementary Section 6.}
\end{figure*}
\begin{figure*}[t]
	\includegraphics[width=\textwidth]{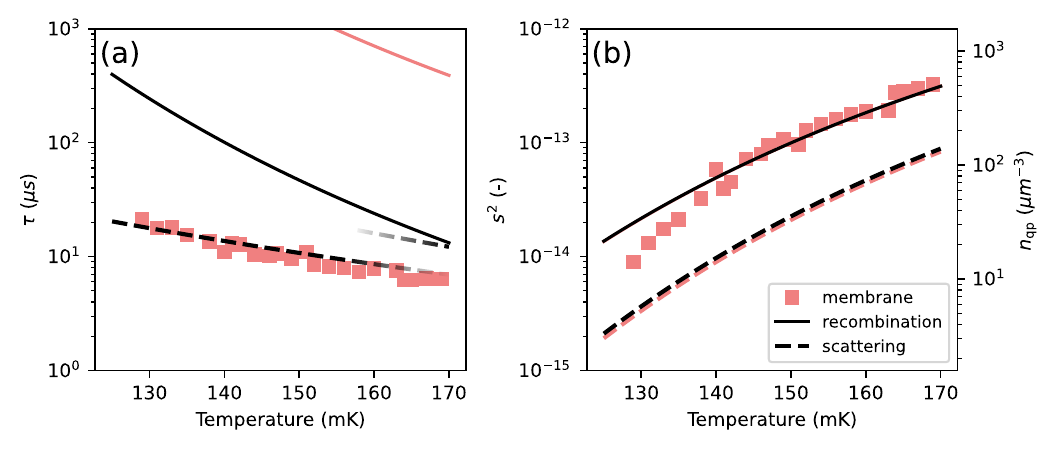}
	\caption{\label{fig:s1results}\textbf{Fluctuation measurement of $\sigma_1$.} Quasiparticle relaxation time (a) and the variance (b) are obtained the same way as in \cref{fig:results}, but from the fluctuations of the dissipative part of the complex conductivity, $\delta\sigma_1$. Error bars indicating statistical fit errors are smaller than the data points. The quasiparticle signal in $\sigma_1$ is a few times lower than in $\sigma_2$ \cite{Gao2008}, which results in a smaller usable temperature range. For the substrate resonator we did not observe any Lorentzian signatures in the PSDs of $\sigma_1$.}
\end{figure*}
\begin{figure*}[t]
	\includegraphics[width=\textwidth]{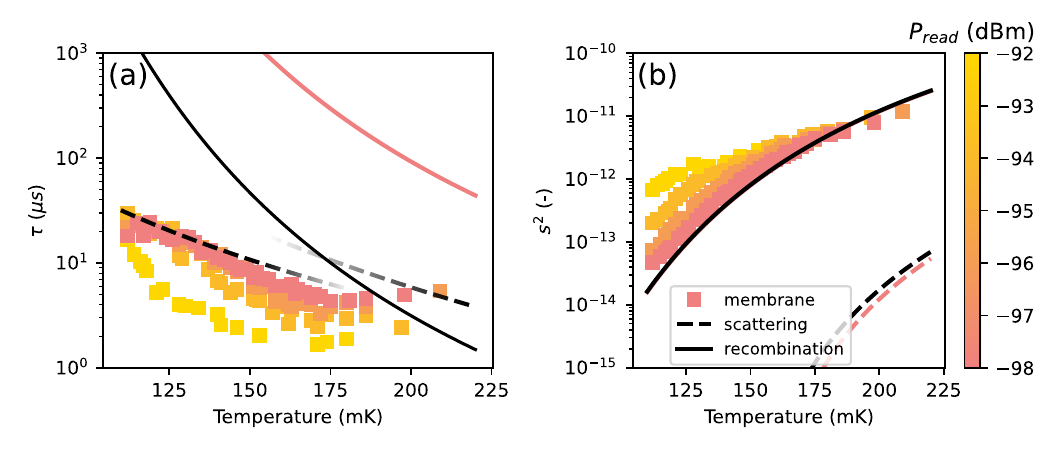}
	\caption{\label{fig:Pread}\textbf{Read-out power dependence of the relaxation time and variance.} (a) Relaxation time versus temperature for the membrane resonator at different on-chip read-out powers, indicated by the colors. The red data points and theory lines are the same as in \cref{fig:results}. (b) Corresponding variances extracted from the same fits to $\sigma_2$-fluctuations. Error bars indicating the statistical fit errors are smaller than the data points, for both panels. A higher read-out power, leads to a larger variance and smaller lifetime, which is consistent with microwave power induced excess quasiparticles \cite{deVisser2014b}. To limit these effects, we set the read-out power to the minimal value in this figure (red data set) and obtain the data in \cref{fig:results}. See Supplementary Section 7 for an estimation of the read-out power effects at this lowest read-out power.}
\end{figure*}
Before each fluctuation measurement, we sweep the probe frequency to find the resonance frequency, $f_0$, and calibrate the complex forward transmission to an amplitude, $\delta A$, and phase, $\theta$, with respect to the resonance circle \cite{deVisser2014}. We set the microwave power of the on-chip probe tone relatively low at $-98~\text{dBm}$ to limit non-equilibrium effects such as quasiparticle redistribution \cite{deVisser2014b} (see \cref{fig:Pread}). We measure $\theta$ during 40 s at 50 kHz sampling frequency and 1 s at 1 MHz. We disregard parts of the time traces that contain pulses from cosmic rays and calculate the Power Spectral Densities (PSDs) \cite{deRooij2021} and stitch the two PSDs from the 50 kHz and 1 MHz data at 20 kHz and downsample the spectrum to 30 points per decade to obtain $S_{\theta}(\omega)$ for a single bath temperature. The fluctuations in $\sigma_2$, $f_0$ and $\theta$ are related via, 
\begin{align}
\begin{split}
		S_{\delta \sigma_2/|\sigma|}(\omega) &= \left(\frac{4}{\alpha_\text{k} \beta }\right)^2S_{\delta f/f_0}(\omega) \\
		&=  \left(\frac{1}{\alpha_\text{k} \beta Q}\right)^2S_{\theta}(\omega).
\end{split}
\end{align}
Here, $\sigma=\sigma_1-i\sigma_2$ the complex conductivity and $|\sigma|$ is the absolute, mean, value at the set bath temperature. At low temperatures ($k_\text{B}T\ll\Delta_0$) $|\sigma|\approx\sigma_2$. $Q = (1/Q_\text{i} + 1/Q_\text{c})^{-1}$ is the loaded quality factor and $\alpha_\text{k}=L_\text{k}/L_{\text{tot}}$ is the kinetic inductance fraction of the $\beta$-Ta volume with respect to the entire resonator. $\beta$ is a correction factor for the film thickness, which we set to 2 since we are in the thin film limit ($\lambda\gg d$, with $\lambda\approx 1.6~\mu \text{m}$, the penetration depth) \cite{deVisser2014}. \\
$Q$ and $\alpha_\text{k}$ are measured in a separate measurement where we sweep the probe frequency to get the resonance curve at each bath temperature. We fit a Lorentzian resonance dip to those curves and extract $f_0$, $Q_\text{i}$ and $Q_\text{c}$ versus temperature. From $f_0(T)$ we determine $\alpha_\text{k}$ \cite{deVisser2014}, which in our case is $\alpha_\text{k}\approx 0.44$.\\
We disregard the 50 Hz, amplifier and $1/f$ noise contributions in the PSD, see Supplementary Section 4. After that, we fit a Lorentzian spectra to $S_{\delta \sigma_2/|\sigma|}(\omega)$ via,
\begin{equation}\label{eq:Sfit}
	S_{\text{fit}}(\omega)=\frac{4s^2\tau}{1+(\omega\tau)^2},
\end{equation}
to extract the variance $s^2$ and relaxation time $\tau$ from the fluctuations. \\
The resonance frequencies and quality factors result in a resonator bandwidth of $\Delta f = f_0/(2Q) \approx 0.3~\text{MHz}$ (or equivalently, a resonator ring time of $\sim 0.6~\mu\text{s}$), which is outside of the frequency range used for the fits. We therefore do not consider effects of the resonator roll-off on the PSDs.\\
The datasets for membrane and substrate have been measured at two different times, in the same setup and experimental conditions.
\paragraph{Data and code availability}
	All data and analysis scripts used in this study have been deposited in the Zenodo database under accession code 13380277: \href{https://zenodo.org/records/13380277}{10.5281/zenodo.13380277}. 
\paragraph{Acknowledgements}
P.J.d.V. and K.K. were supported by the Netherlands Organisation for Scientific Research NWO (Veni Grant No. 639.041.750 and Projectruimte 680-91-127). J.J.A.B. was supported by the European Research Council ERC (Consolidator Grant No. 648135 MOSAIC). 
\paragraph{Author contributions}
P.J.d.V. and S.A.H.d.R. conceived the experiment. S.A.H.d.R. designed and T.C, V.M. and D.J.T fabricated the device. S.A.H.d.R. performed the microwave measurements, which P.J.d.V. supervised. R.F. performed the characterization measurements, which J.A. supervised. S.A.H.d.R. analyzed the data and S.A.H.d.R., R.F., K.K., J.A., J.J.A.B. and P.J.d.V discussed the results. S.A.H.d.R. wrote the manuscript and R.F., K.K., J.J.A.B. and P.J.d.V. reviewed it substantively. The project was supervised by J.J.A.B. and P.J.d.V.
\paragraph{Ethics declarations}
The authors declare to have no competing interest.

\clearpage
\renewcommand{\theequation}{S.\arabic{equation}}
\renewcommand{\thefigure}{S\arabic{figure}}
\renewcommand{\thetable}{S\arabic{table}}
\renewcommand{\thesection}{\arabic{section}}
\renewcommand{\tablename}{Table}
\renewcommand{\figurename}{Fig.}

\title{Supplementary Information for: '\textit{Recombination of localized quasiparticles in disordered superconductors}'}

\maketitle

\onecolumngrid

\section{Relaxation time and variance calculations}\label{sec:tauvar}
Here we describe how the theory lines in \cref{fig:results}(a) and (b) are calculated. The fluctuation spectrum is given by a Lorentzian, $S(\omega)=4s^2\tau/(1+(\omega\tau)^2)$, with $\tau$ the relaxation time and $s^2$ the variance.  We will first go over the relaxation time calculations (\cref{fig:results}(a)), including the phonon trapping effect for the $\beta$-Ta on substrate and membrane, and then explain how we calculated the variances (\cref{fig:results}(b)).

\subsection*{Relaxation time} 
\subsubsection*{Electron-phonon interaction}
First, we recall that the $\beta$-Ta film is disordered with respect to electron-phonon interactions, i.e. $q(\Omega)l<1$ with $q(\Omega)$ the phonon wave number and $l$ the electronic mean free path, for both thermal ($\Omega=k_\text{B}T$) and recombination ($\Omega=2\Delta_0$) phonons. $k_\text{B}$ is the Boltzmann constant, $T$ is the bath temperature and $2\Delta_0$ is the superconducting gap energy at low temperatures and when there is no disorder present.\\
Second, the film is 2D for $T<165~\text{mK}$ thermal phonons ($qd<1/2$, with $d$ the film thickness) and 3D for higher temperatures and for recombination phonons. We take the values for quasiparticle energy $\Delta_0$, by which we assume that the quasiparticles are in thermal equilibrium and are all relaxed to the gap energy.
\paragraph*{Recombination} 
The 3D case of the recombination time for disordered superconductors is given in Ref. \cite{Reizer1986}, Eq. (61), 
\begin{equation}\label{eq:trec_e-ph}
	\tau_\text{rec}^\text{e-ph}(T)=\frac{\tau_\text{s}(T_\text{c})}{4\pi^{5/2}}\left(\frac{k_\text{B}T_\text{c}}{2\Delta_0}\right)^{7/2}\sqrt{\frac{T_\text{c}}{T}}e^{\Delta_0/k_\text{B}T},
\end{equation}
where $\tau_\text{s}(T_\text{c})$ is the electron-phonon scattering time in the normal metal at $T=T_\text{c}$. This is given in Eq. (31) of Ref. \cite{Reizer1986} or Eq. (51) of Ref. \cite{Sergeev2000} with $k=1$: 
\begin{equation}\label{eq:taus}
	\tau_\text{s}(T_\text{c})=\frac{5\hbar^4}{\pi^4}\frac{(k_\text{F} c_\text{L})^3}{(k_\text{F} l) (k_\text{B} T_\text{c})^4 \beta_\text{L} \left(1 + \frac{3}{2} \left( \frac{c_\text{L}}{c_\text{T}} \right)^5 \right)},
\end{equation}
with, 
\begin{equation}
	\label{eq:betaL}
	\beta_\text{L} = \left( \frac{2 E_\text{F}}{3} \right)^2 \frac{N_0}{2 \hat{\rho} c_\text{L}^2}.
\end{equation}
Here, $\hbar$ is the reduced Planck constant, $E_\text{F}$ is the Fermi energy, $k_\text{F}$ is the Fermi wavenumber, $c_\text{L}$ and $c_\text{T}$ are the longitudinal and transverse sound velocities, $\hat{\rho}$ is the mass density, $T_\text{c}$ is the critical temperature and $N_0$ is the single spin density of states at the Fermi energy. We use the parameters from \cref{tab:params} and calculate $N_0$ from the Einstein relation, $N_0=1/(2e^2\rho_\text{N} D)$, with $e$ the electronic charge, $\rho_\text{N}$ the normal state resistivity and $D$ the diffusion constant. For $\Delta_0$ we take the zero temperature BCS value, $\Delta_0=1.76k_\text{B}T_\text{c}$, which is correct up to 0.1\%, since we measure at $T<T_\text{c}/4$. We assume the free electron model, so we take $k_\text{F}=(3\pi^2n_e)^{1/3}$, where $n_e$ is the charge density, and $E_\text{F}=k_\text{F}^3/(4 \pi^2 N_0)$. We determine $n_e$ from a Hall resistance measurement and $D$ from an upper critical field measurement, which are explained in the section \textit{Characterization of the $\beta$-Ta film}.

\paragraph*{Scattering} 
For scattering, we need to distinguish between the 2D and 3D phonon case. The 2D case (for $T>165~\text{mK}$) can be found in Ref. \cite{Devereaux1991a}, Eq. (4.8a), bottom row: 
\begin{equation}\label{eq:tscat_e-ph2D}
	\tau_\text{scat}^\text{e-ph,~2D}=\frac{16}{\sqrt{2}}\frac{Z'}{\Gamma(7/2)\zeta(7/2)}\left(\frac{E_\text{F}}{\Delta_0}\right)^2\frac{ c_\text{T}\hat{\rho} d}{\Delta_0 k_\text{F}^3}\frac{\rho_\text{N} e^2}{d\hbar}\frac{c_\text{T}^3}{v_\text{F}^3}\frac{1}{1+(c_\text{T}/c_\text{L})^4}\left(\frac{\Delta_0}{k_\text{B}T}\right)^{7/2},
\end{equation}
where $Z'$ is the Eliashberg renormalization constant for the 3D clean limit, $Z'=1+N_0 (k_\text{B} T_\text{D})^2 v_\text{F}^2 / (36\hat{\rho} c_\text{L}^4)$ \cite{Keck1976}, with $T_\text{D}$ the Debye temperature, see \cref{tab:params}. $v_\text{F}$ is the Fermi velocity, $\Gamma$ is the gamma function and $\zeta$ is the Riemann zeta function.\\
The 3D case can be found in Ref. \cite{Reizer1986}, Eq. (62), 
\begin{equation}\label{eq:tscat_e-ph3D}
	\tau_\text{scat}^\text{e-ph,~3D} = \frac{\tau_\text{s}(T_\text{c})}{4\pi^2\Gamma(9/2)\zeta(9/2)}\sqrt{\frac{2\Delta_0}{k_\text{B}T_\text{c}}}\left({\frac{T_\text{c}}{T}}\right)^{9/2},
\end{equation}  
with $\tau_\text{s}(T_\text{c})$ given by \cref{eq:taus}.

\subsubsection*{Electron-electron interaction}
Electron-electron interactions can be enhanced in disordered metals \cite{Schmid1974}. However, in a superconductor the normal state charge carrier density is exponentially suppressed, such that the electron-phonon interaction become more favorable. To quantify that statement, we calculate the electron-electron interactions times here as well.\\
Our $\beta$-Ta film is in the dirty (or impure) regime since, $\xi/l=62$. $l$ is $0.30~\text{nm}$ and $\xi$ is the dirty limit coherence length, $18~\text{nm}$. The electron-electron rates in the dirty regime are calculated in Ref. \cite{Reizer2000}, both for 2D and 3D. This is an extension of the theory in Ref. \cite{Devereaux1991a} to include gapless collective phase mode in 2D, which occurs when $k_\text{F}\xi \lesssim 1$. This collective mode has a power law temperature dependence with $n=7/2$. In our case, $k_\text{F}\xi \approx 262$ so we are not in that regime. Besides that, $\xi/d=0.47$, so we are not in the 2D but 3D regime. Therefore, the observed power law in \cref{fig:results} cannot be this gapless phase mode.\\
All other interaction channels for the electron-electron interaction are gapped and have a exponential temperature dependence. From Ref. \cite{Reizer2000}, Eq. (48), the 3D case, we have, 
\begin{equation}\label{eq:trec_e-e}
	\tau_\text{rec}^\text{e-e} = \frac{(\hbar\sqrt{2})^{5/2} (2\pi)^2 N_0 D}{k_\text{B}T}\sqrt{\frac{D}{\Delta_0}} \frac{e^{2\Delta_0/k_\text{B}T}}{\left(1 + (V_\text{sc}N_0)^2\right)}.
\end{equation}
$V_\text{sc}$ is the BCS attractive potential energy. We calculate this via the BCS gap equation at zero temperature, $1/V_\text{sc}=N_0\int_{\Delta_0}^{k_\text{B}T_\text{D}} dE/\sqrt{E^2-\Delta_0^2}=1.2\times10^{5}~\mu\text{eV}^{-1}\mu\text{m}^{-3}$.\\
For scattering, we have from Ref. \cite{Reizer2000}, Eq. (47),
\begin{equation}\label{eq:tscat_e-e}
	\tau_\text{scat}^\text{e-e} = \frac{(\pi\hbar)^{5/2} N_0 D}{6\Delta_0}\sqrt{\frac{D}{k_\text{B}T}} e^{4\Delta_0/5k_\text{B}T}.
\end{equation}

\begin{figure*}[t]
	\includegraphics[width=.6\textwidth]{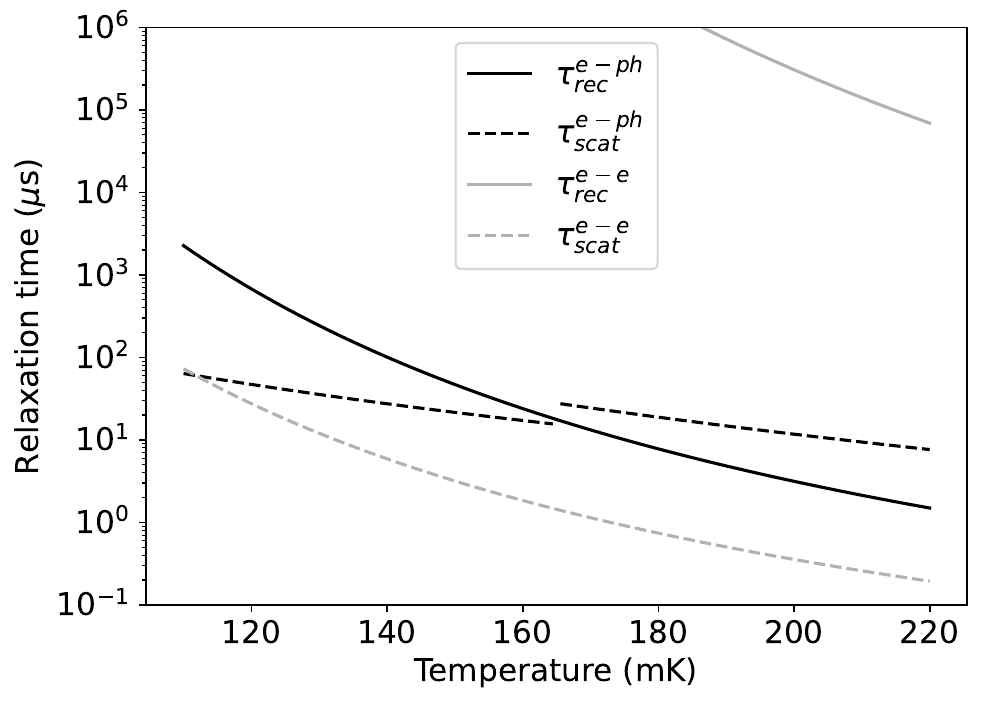}
	\caption{\label{fig:lifetimes} Electron-phonon and electron-electron relaxation rates from \cref{eq:trec_e-ph,eq:tscat_e-ph2D,eq:tscat_e-ph3D,eq:trec_e-e,eq:tscat_e-e}. The temperature range is chosen to match our measurement regime. The electron-phonon times (black solid and dashed lines) are the same as in \cref{fig:results}(a).} 
\end{figure*}

In \cref{fig:lifetimes} the recombination and scattering lifetimes for electron-phonon and electron-electron interactions are shown. $\tau_\text{rec}^\text{e-e}$ is an order of magnitude longer than $\tau_\text{rec}^\text{e-ph}$ due to the additional factor $e^{\Delta_0/k_\text{B}T}$, which reflects the 3 particle nature of electron-electron recombination instead of 2 quasiparticle electron-phonon recombination.\\
The electron-electron scattering time is the shortest in this temperature regime. However, this process conserves quasiparticle number and does not change the total energy stored in the quasiparticle system. It only changes the distribution function to the thermal Fermi-Dirac distribution on the time scale of $\tau_\text{scat}^\text{e-e}$. As we calculated for electron-phonon scattering, the variance on the kinetic inductance for changes in distribution function are very low (see \cref{fig:results}(b)) and this is not visible in our experiment.

\subsubsection*{Phonon trapping factor}
The phonons that are emitted during a recombination event can break other Cooper-pairs, which enhances the measured bulk quasiparticle lifetime. This effect is captured by the phonon trapping factor, $(1+\tau_\text{esc}/\tau_\text{pb})$ in \cref{eq:taurecstar} \cite{Rothwarf1967, Wilson2004}. We experimentally tune $\tau_\text{esc}$ by measuring the $\beta$-Ta on substrate and on a thin membrane. We will first estimate $\tau_\text{esc}$ for $\beta$-Ta on a SiN substrate and then estimate the enhanced phonon trapping by the membrane.\\
We assume that phonon scattering on the film boundaries is diffusive. The SiN is etched by a reactive ion etch when patterning the NbTiN before deposition of $\beta$-Ta, which makes the interface rough. We only take transverse phonons into account (since $(c_\text{T}/c_\text{L})^3 \ll 1$) and neglect phonon loss in the bulk. We then use Eq. (51) of Ref. \cite{Eisenmenger1976}, 
\begin{equation}\label{eq:phtrf}
	(1 + \tau_\text{esc}/\tau_\text{pb}) = \frac{4d}{\Lambda_T}\frac{1 - (1 - \eta_T)e^{-2d/\Lambda_T}}{\eta_T (1 - e^{-2d/\Lambda_T})}.
\end{equation}
$\Lambda_T$ is the transverse phonon mean free path against pair breaking, which is given by, $\Lambda_T=c_\text{T}\tau_\text{pb}^T$, with $c_\text{T}$ from \cref{tab:params} and $\tau_\text{pb}^T$ the pair breaking time for transverse phonons. We estimate $\tau_\text{pb}^T$ as the inelastic scattering time of phonons in a metal \cite{Pippard1955,Kittel1987}, 
\begin{equation}
	\frac{1}{\tau_\text{pb}^T}=\frac{n_em^*}{\hat{\rho} \tau}\left(\frac{1}{\zeta} - 1\right),
\end{equation}
with $\zeta=3[(1+(ql)^2)\tan^{-1}ql-ql]/2(ql)^3$, where $q=2\Delta_0/(\hbar c_\text{T})$ is the recombination phonon wave number. $n$ is the charge density, $m^*$ is the effective electron mass, $m^*=2\hbar^2 N_0 \pi^2/k_\text{F}$ and $\tau_\text{e}$ is the Drude elastic scattering time, $\tau_\text{e}=l/v_\text{F}$. This gives $\tau_\text{pb}^T=37~\text{ns}$, which is relatively long due to disorder, $ql<<1$ \cite{Pippard1955}. Thus, $\Lambda_T=65~\mu\text{m}$ is also large. For example, $\tau_\text{pb}^T$ in Al is 0.12 ns and $\Lambda_T$ is 0.39 $\mu$m.\\
$\eta_T$ in \cref{eq:phtrf} is the transmission efficiency for transverse phonons. We it calculate via the acoustic mismatch model from Ref. \cite{Kaplan1979}. The code to calculate the phonon transparencies for an arbitrary interface is available at Zenodo (\href{https://zenodo.org/records/13380277}{10.5281/zenodo.13380277}).\\
We use the values from \cref{tab:params} for the $\beta$-Ta phonon properties and $\hat{\rho}=3.1~\text{g/cm\textsuperscript{3}}$, $c_\text{T} = 6.2~\text{km/s}$ and $c_\text{L} = 10.3~\text{km/s}$ for SiN \cite{Kuhn2004, Petersen1982}. That gives $\eta_T=0.054$, which results in a phonon trapping factor, from \cref{eq:phtrf}, of $(1+\tau_\text{esc}/\tau_\text{pb})=2.0$.

\begin{figure*}[t]
	\includegraphics[width=.8\textwidth]{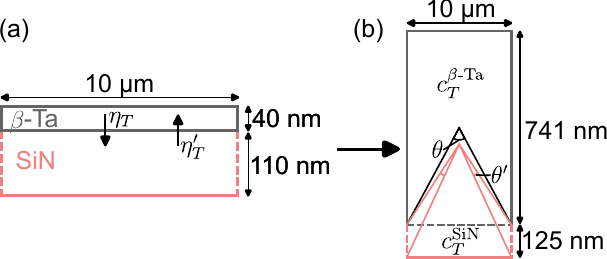}
	\caption{\label{fig:phonontrapping} Illustration of the geometrical calculation to estimate the effect of the membrane on the phonon escape time, $\tau_\text{esc}$. (a): Situation of the $\beta$-Ta film (grey) on the SiN membrane (pink). The physical thicknesses are indicated. The dashed lines indicate that it is not a physical boundary of the material, but only of the volume considered. $\eta_T$ is the transverse phonon transparency from $\beta$-Ta to SiN and $\eta_T'$ vice versa. (b): Effective film thicknesses $d/\eta_T$ and $d_\text{SiN}/\eta_T'$, with the phonon escape angles for the substrate case ($\theta$) and the membrane case ($2\theta'$) as indicated.}
\end{figure*}

To estimate the effect of the membrane, we use a geometrical calculation as shown in \cref{fig:phonontrapping}. The $2\Delta_0$-phonon wavelength is $27~\text{nm}$, which is smaller but close to $2d=80~\text{nm}$, so a purely geometrical calculation is somewhat justified. We will come back to this point at the end of this section.\\
We calculate the transverse phonon transparency from SiN to $\beta$-Ta in the same way as explained above. This results in $\eta_T'=0.80$. We use the out- and in-going transparencies ($\eta_T$ and $\eta_T'$, respectively) to scale the thickness of the two layers, see \cref{fig:phonontrapping}(b). With these effective thicknesses, we take the phonon transparencies in the new geometry, \cref{fig:phonontrapping}(b), to be 1, so we can consider it as one geometrical volume. We assume scattering on the sides is diffusive, such that all the angle dependencies are averaged out. We therefore only consider the escape angles as measured from the center of the geometrical volume. This is slightly different for the membrane compared to the substrate case, see \cref{fig:phonontrapping}(b).\\
For the substrate case we assume a phonon is lost when it enters the SiN. This results in the escape angle $\theta$ in \cref{fig:phonontrapping}(b). For the membrane case, we consider the phonon to be lost when it leaves the volume beneath the $\beta$-Ta, which results in an escape angle of $2\theta'$ in \cref{fig:phonontrapping}(b). The effect of the membrane is given by the ratio of the these two escape angles, $\theta/(2\theta')$, when we also compensate for the time that the phonon is in the SiN membrane via the effective thicknesses and transverse sound velocities, 
\begin{equation}
	\frac{\tau_\text{esc}^\text{mem}}{\tau_\text{esc}^\text{sub}} = \frac{\theta}{2\theta'}\frac{d/(\eta_Tc_\text{T}^{\beta\text{-Ta}})}{d_{\text{SiN}}/(\eta_T' c_\text{T}^\text{SiN}})+d/(\eta_T c_\text{T}^{\beta\text{-Ta}}) = 60\times0.96=58.
\end{equation}
This gives a phonon trapping factor of $(1 + \tau_\text{esc}^\text{mem}/\tau_\text{pb})=59$. The black line in \cref{fig:results}(c) is given by $\tau_\text{rec}^*=\tau_\text{rec}^\text{e-ph}(1+\tau_\text{esc}/\tau_\text{pb})/2$ (\cref{eq:taurecstar}), where $\tau_\text{rec}^\text{e-ph}$ is the solid black line in \cref{fig:lifetimes} and $(1+\tau_\text{esc}/\tau_\text{pb})=2.0$. The red line in \cref{fig:results} is the same, but with $(1+\tau_\text{esc}/\tau_\text{pb})=59$.\\
As a check, we perform the exact same calculation for Al on SiN and compare it with Ref. \cite{deRooij2021}. We come to a phonon trapping factor on SiN substrate of 3.2 and on membrane 17.6. So, a factor 5.5 longer lifetimes are expected. The measurement in Ref. \cite{deRooij2021} shows an increase in lifetime of a factor 16. Therefore, the assumption of a purely geometric effect of the membrane has limited applicability. For Al, the $2\Delta_0$-phonon wavelength is 37 nm, while the film thickness is 50 nm. For higher energetic phonons (i.e. shorter wavelengths) generated by single photon absorption events, such a geometric model is consistent with the data \cite{deVisser2021}. We therefore expect that the low energetic $2\Delta_0$-phonons are trapped more effectively by the membrane than the geometrical calculation predicts. Therefore, we can interpret the above geometric calculation as minimum for the phonon trapping effect of the membrane.

\subsection*{Variance}
The theory lines for the variance in \cref{fig:results}(d) are calculated as,
\begin{equation}\label{eq:s2varresp}
	s^2=\left<\delta (\sigma_2/|\sigma|)^2\right>=\left<\delta(k_\text{B}T)^2\right>\left(\frac{d(\sigma_2/|\sigma|)}{dk_\text{B}T}\right)^2,
\end{equation}
where the first factor is the variance of the thermal fluctuations and the second is the responsivity to temperature changes. We will calculate these factors for scattering and recombination separately. 
\subsubsection*{Thermal fluctuations}
In equilibrium, thermal fluctuations have a variance of $\left<\delta(k_\text{B}T)^2\right>=k_\text{B}^3T^2/C$, where $C=dU/dT$ is the heat capacity, with $U$ the internal energy \cite{Landau2013}.

\paragraph*{Recombination}
For recombination, $C$ is given by the change in quasiparticle number, $\delta N_\text{qp}$, with energy $\Delta_0$. That gives a heat capacity \cite{Thomas2015},
\begin{align}\label{eq:s2rec}
	\begin{split}
		C_\text{qp}=\frac{dU}{dT}&=\frac{d}{dT}\left(4N_0 V \int_{\Delta_0}^\infty\frac{E^2f(E; k_\text{B}T)}{\sqrt{E^2-\Delta_0^2}}dE\right)\\
		&\approx \frac{d}{dT}\left(2N_0 V \Delta_0^2 \left[K_0(\Delta_0/k_\text{B}T) + K_2(\Delta_0/k_\text{B}T\right]\right)\\
		&\approx \frac{d}{dT}\left(\Delta_0 N_\text{qp}\right)\\
		&\approx \frac{N_\text{qp}\Delta_0^2}{k_\text{B}T^2}.
	\end{split}
\end{align}
$K_n(x)$ is the $n$-th modified Bessel function of the second kind. The first approximation is valid for $f(E; k_\text{B}T)\approx e^{-E/k_\text{B}T}$. The other two are valid for $k_\text{B}T \ll \Delta_0$, which automatically satisfies the first. 
From this we see that to first order in $k_\text{B}T/\Delta_0$, the internal energy is changed mainly by the change in quasiparticle number and not by the change in distribution function.\\ 
If we use this heat capacitance, we come to a variance of 
\begin{equation}\label{eq:varrec}
	\left<\delta(k_\text{B}T)^2\right>_\text{rec}=\frac{(k_\text{B}T)^4}{N_\text{qp}\Delta_0^2}.
\end{equation}
If we multiply this variance with $(dN_\text{qp}/dk_\text{B}T)^2\approx (N_\text{qp}\Delta_0)^2/(k_\text{B}T)^4$, we get $N_\text{qp}$, which is the variance of quasiparticle number fluctuations ($\left<\delta N_\text{qp}^2\right>=N_\text{qp}$). Thus, this is the variance of the temperature fluctuations due to recombination. 
\paragraph*{Scattering}
For scattering, we can do something similar as Ref. \cite{Gao2008}: we implicitly introduce a chemical potential $\mu^*$ to keep $N_\text{qp}$ constant, while changing $k_\text{B}T$. In this case, we need the higher orders in the Bessel functions $K_0(\Delta_0/k_\text{B}T)$ and $K_2(\Delta_0/k_\text{B}T)$, to get the first order in explicit temperature dependence of $U$. So, 
\begin{align}
	\begin{split}
	U &= N_0 V \Delta_0\sqrt{2\pi k_\text{B}T\Delta_0}e^{-\frac{\Delta_0}{k_\text{B}T}}\left(1+\frac{k_\text{B}T}{\Delta_0}\right)\\
&=\Delta_0 N_\text{qp}\left(1+\frac{k_\text{B}T}{\Delta_0}\right).
	\end{split}
\end{align}

Now, 
\begin{equation}
	C_\text{rec}=\frac{\partial U}{\partial N_\text{qp}}\frac{dN_\text{qp}}{dT}=\frac{N_\text{qp}\Delta_0^2}{k_\text{B}T^2},
\end{equation} 
and 
\begin{equation}
	C_\text{scat}=\frac{\partial U}{\partial T}=k_\text{B} N_\text{qp}.
\end{equation}
The total variance can be written, 
\begin{equation}\label{eq:thmvar}
	\left<\delta(k_\text{B}T)^2\right>=\frac{k_\text{B}^3T^2}{C_\text{rec}+C_\text{scat}}\approx\frac{(k_\text{B}T)^4}{N_\text{qp}\Delta_0^2}\left[1-\left(\frac{k_\text{B}T}{\Delta_0}\right)^2\right],
\end{equation}
where the approximation is valid for $k_\text{B}T\ll \Delta_0$. This shows that scattering only gives a small negative correction to the variance of the temperature fluctuations. It is a negative correction, as the change in $k_\text{B}T$ enhances the heat capacity.

If the number of quasiparticles stays constant, $C_\text{rec}=0$ and we would be left with, 
\begin{equation}\label{eq:varscat}
	\left<\delta(k_\text{B}T)^2\right>_\text{scat}=\frac{k_\text{B}^3T^2}{C_\text{scat}}=\frac{(k_\text{B}T)^2}{N_\text{qp}}. 
\end{equation}
This is larger than $\left<\delta(k_\text{B}T)^2\right>_\text{rec}$ by a factor $(\Delta_0/k_\text{B}T)^2$, which is $\sim10^{2}$ in our measurement, see \cref{fig:varresp}(a). We use \cref{eq:varscat} for the variance of scattering events. It is however only valid when the number of quasiparticles stays constant. This is true when $\tau_\text{scat}\ll\tau_\text{rec}$, which is valid for $T\lesssim 130~\text{mK}$ in our case, see \cref{fig:lifetimes}. When this condition is not satisfied, the variance will be governed by recombination events. Therefore, \cref{eq:varscat} can be viewed as an upper limit for the complete measurement range.

\subsubsection*{Responsivities}
\begin{figure*}[t]
	\includegraphics[width=.9\textwidth]{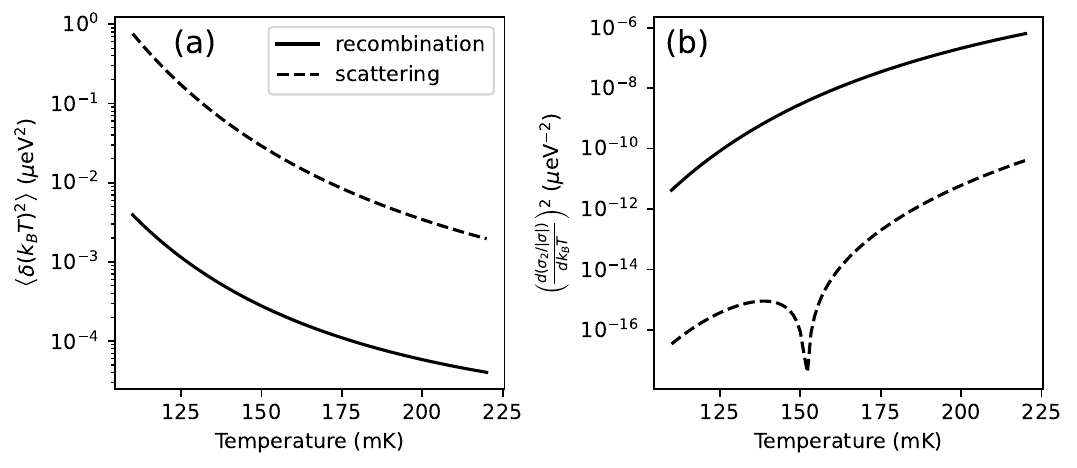}
	\caption{\label{fig:varresp} Contributions to the variance calculations. (a): Thermal fluctuation variance for recombination, \cref{eq:varrec}, and scattering, \cref{eq:varscat}. (b): Temperature responsivity of $\sigma_2/|\sigma|$ from \cref{eq:respscat}. The legend is applicable to (a) and (b). For the calculations we set $\omega_0/2\pi=5~\text{GHz}$ and use the properties of $\beta$-Ta from \cref{tab:params}. The multiplication of the curves in (a) and (b) give the theory lines in \cref{fig:results}(b).}
\end{figure*}
The second factor in \cref{eq:s2varresp} is the temperature responsivity of $\sigma_2/|\sigma|$ squared. We calculate this factor for recombination and scattering separately. We use the methods of Ref. \cite{Gao2008} to calculate these responsivities from the Mattis-Bardeen equations \cite{Mattis1958}. For completeness, we show the calculations for both $\sigma_1$ and $\sigma_2$. 

For recombination, we assume that the temperature is constant, while the number of quasiparticle changes with the use of a chemical potential. We can use Eqs. (18) and (19) from Ref. \cite{Gao2008} and multiply by 
\begin{equation}
	\frac{dn_\text{qp}}{dk_\text{B}T}\approx\frac{n_\text{qp}\Delta_0}{(k_\text{B}T)^2}.
\end{equation}
The approximation is valid for low temperatures, $k_\text{B}T \ll \Delta_0$. This results in,
\begin{align}\label{eq:resprec}
	\begin{split}
		\left.\frac{d\sigma_1}{dk_\text{B}T}\right\rvert_\text{rec} &= \frac{\sigma_\text{n} n_\text{qp}}{N_0\hbar\omega}\sqrt{\frac{2\Delta_0^3}{\pi (k_\text{B}T)^5}}\left[\sinh(\chi)K_0(\chi)\right]\\
		\left.\frac{d\sigma_2}{dk_\text{B}T}\right\rvert_\text{rec} &= -\frac{\sigma_\text{n}n_\text{qp}}{N_0\hbar\omega}\sqrt{\frac{\pi\Delta_0^3}{2 (k_\text{B}T)^5}}\left[\sqrt{\frac{\pi k_\text{B}T}{2\Delta_0}}+e^{-\chi}I_0(\chi)\right],
	\end{split}
\end{align}
with, $\chi={\hbar\omega_0}/{2k_\text{B}T}$. $I_0(\chi)$ is the zero-order modified Bessels function of the first kind. $\omega_0$ is the angular resonance frequency of the resonator.

For scattering, we assume that the number of quasiparticles is constant, but $k_\text{B}T$ changes. We take Eqs. (16) and (17) from Ref. \cite{Gao2008} and differentiate with respect to $k_\text{B}T$ to find:
\begin{align}\label{eq:respscat}
	\begin{split}
	\left.\frac{d\sigma_1}{dk_\text{B}T}\right\rvert_\text{scat} &= \frac{\sigma_\text{n} n_\text{qp}}{N_0\hbar\omega}\sqrt{\frac{\Delta_0}{2\pi (k_\text{B}T)^3}}\left[2\chi\left(\sinh(\chi) K_1(\chi)-\cosh(\chi)K_0(\chi)\right)-\sinh(\chi)K_0(\chi)\right]\\
	\left.\frac{d\sigma_2}{dk_\text{B}T}\right\rvert_\text{scat} &= \frac{\sigma_\text{n} n_\text{qp} }{N_0\hbar\omega}\sqrt{\frac{\pi\Delta_0}{8 (k_\text{B}T)^3}}\left[e^{-\chi}\left(I_0(\chi)+2\chi\left(I_1(\chi)-I_0(\chi)\right)\right)\right].
	\end{split}
\end{align}

We divide \cref{eq:resprec,eq:respscat} by $|\sigma(T)|=|\sigma_1-i\sigma_2|$, which we calculate $\sigma(T)$ from the Mattis-Bardeen equations \cite{Mattis1958}. When we square the result, we get for $\sigma_2/|\sigma|$ the values plotted in \cref{fig:varresp}(b), for $\omega_0/2\pi=5~\text{GHz}$ and the material parameters for $\beta$-Ta from \cref{tab:params}.\\
The responsivity for scattering events is $\sim5$ orders of magnitude lower than for recombination. This can also be seen from the curves in Fig. 1 of Ref. \cite{Gao2008}: to first order, the scattering responsivity of $\sigma_1$ and $\sigma_2$ is equal to the difference of the thermal and excess quasiparticle responsivity curves. This is very small, but grows for higher temperatures.\\
The dip around 150 mK in \cref{fig:varresp}(b) is at the point where $\chi\approx0.79$, or $k_\text{B}T \approx 0.63\hbar\omega_0$. At that value of $\chi$, the term in square brackets in the lower equations of \cref{eq:respscat} vanishes. \\
The curves in \cref{fig:results}(b) are the multiplication of the curves in \cref{fig:varresp}(a) and (b), but with the actual resonance frequencies. The variance of thermal fluctuations is $\sim 2$ order of magnitude higher but the responsivity is $\sim5$ orders of magnitude lower. Therefore, the variance of $\sigma_2$-fluctuations is $\sim 3$ orders of magnitude lower for scattering events, which is shown in \cref{fig:results}(b).\\
As the substrate and membrane resonator have a slightly different resonance frequency the dip in \cref{fig:varresp}(b) is at a different location. That results in the difference for the variance of $\sigma_2$-fluctuations for scattering events in \cref{fig:results}(b).

\section{Characterization of the $\beta$-T\lowercase{a} film}
\begin{figure*}[t]
	\includegraphics[width=.9\textwidth]{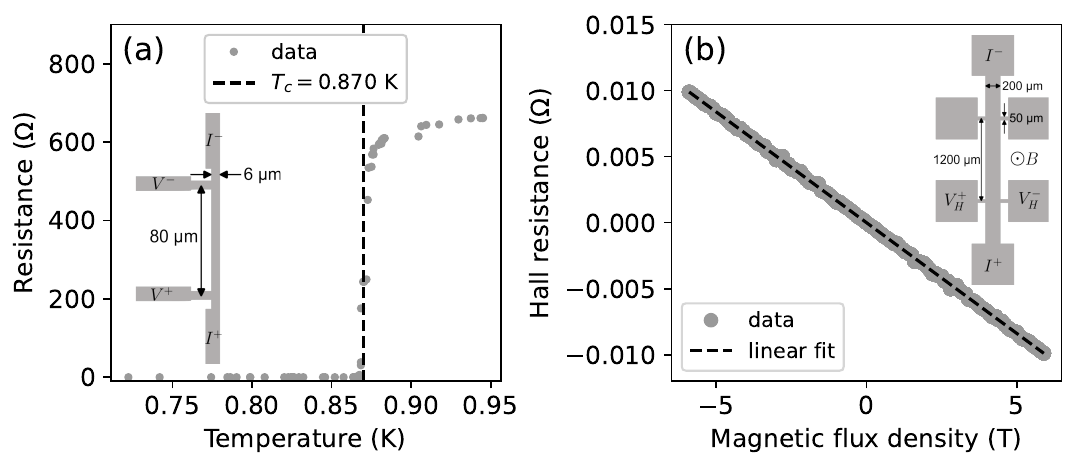}
	\caption{\label{fig:hallres} Transport measurements characterizing the $\beta$-Ta film. (a): Resistance measurement versus temperature in the absence of magnetic field. The dashed line give the critical temperature we use in the analysis. The inset shows the geometry which is used for the measurement. (b): Anti-symmetrized Hall resistance ($V_H/I$) measurement, performed at $1~\text{K}$. The standard deviation for each data point is smaller than the marker size, for both panels. The dashed line gives a linear fit with a slope of  $dR_H/dB=1.68\times10^{-3}~\Omega/\text{T}$. The inset shows the Hall bar geometry and contacts used for the measurement. The measurement in (a) is performed with a different setup than the measurement in (b). The devices are from the same wafer as the devices shown in \cref{fig:results}(a). }
\end{figure*}

We characterize the $\beta$-Ta film in three ways: (1) a resistance versus temperature measurement; (2) a Hall resistance versus magnetic field and (3) a upper critical field measurement versus temperature. The first is measured in the same setup as the measurements presented in the main text, but at a different, thermally weakly coupled, stage. The resistance measurements are done with a Lakeshore model 372 AC resistance bridge.\\
The Hall resistance and upper critical field measurement are performed in a different setup, which includes a superconducting coil to apply a magnetic field up to 6 T, which is orientated in the out-of-plane direction with respect to the $\beta$-Ta film. In this setup, the resistance measurements are performed with a SynkTek MCL1-540 lock-in amplifier. \\
The 4-probe and Hall bar structures used for these measurements are fabricated on the same wafer as the devices presented in the main text. The geometries are shown in the insets of \cref{fig:hallres}.

\subsection*{Resistance versus temperature}
The resistance measurement is shown in \cref{fig:hallres}(a). From this measurement, we get a critical temperature of $T_\text{c}=0.87~\text{K}$ shown by the dashed line. We choose $T_\text{c}$ as the temperature where the resistance is approximately half the normal state resistance. We measure the normal state resistance to be $R_N = 661~\Omega$ at $1.0~\text{K}$, which equals a normal state resistivity $\rho_\text{N}=206~\mu\Omega\text{cm}$ when taking the geometry in \cref{fig:hallres}(a) into account.
\subsection*{Hall resistance}
The Hall resistance measurement is performed at $1.0~\text{K}$ bath temperature and with the geometry shown in the inset of \cref{fig:hallres}(b). The dashed line shows a linear fit to the data, with $dR_H/dB = 1.68\times10^{-3}~\Omega/\text{T}$. The raw resistance measurement data is anti-symmetrized with respect to the magnetic field ($R_\text{Hall} = (R_T(B) - R_T(-B))/2$) to eliminate of any longitudinal resistance component. From the fitted slope we can find the charge carrier density, 
\begin{equation}
	n_e=\frac{1}{ed}\left(\frac{dR_\text{Hall}}{dB}\right)^{-1} = 9.3\times10^{28}~\text{m}^{-3},
\end{equation}
where $e$ is the electron charge and $d$ is the film thickness. Assuming the free electron model to hold, we can calculate the Fermi wavenumber as, $k_\text{F}=(3\pi^2n_e)^{1/3}=14~\text{nm}^{-1}$. The Ioffe-Regel parameter is given by, $k_\text{F}l=\hbar(3\pi^2)^{2/3}/(n_e^{1/3}e^2\rho_\text{N})=4.2$. This also gives the electron mean free path, $l=0.30~\text{nm}$. 

\subsection*{Upper critical field}
\begin{figure*}[t]
	\includegraphics[width=.9\textwidth]{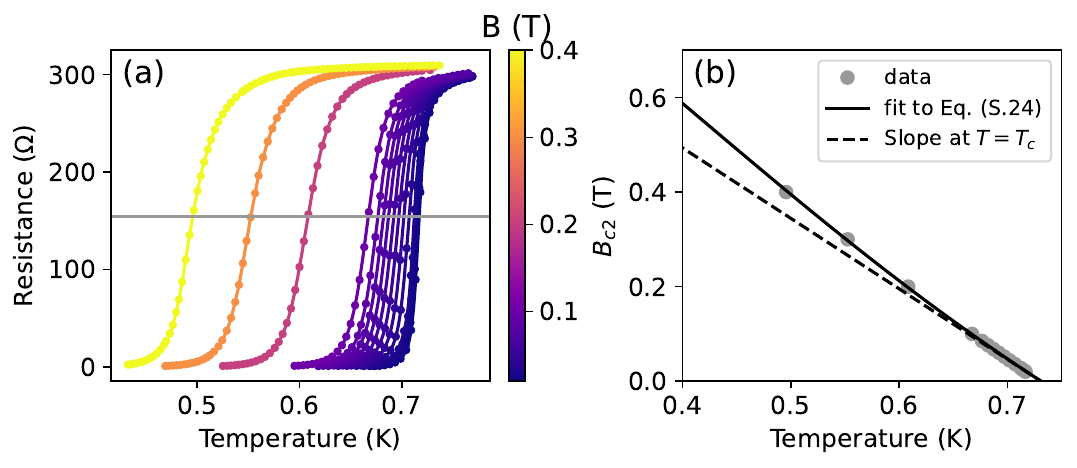}
	\caption{\label{fig:Hc2} Upper critical field measurement versus temperature. (a): Resistance versus temperature at different magnetic flux densities, see the colorbar. The dots are data points and the solid lines are cubic spline fits for each value of $B$. The gray line indicates where we chose to define $B_\text{c2}$, which is arbitrary for our analysis as long as it is in the transition. The measured device is depicted in the inset of \cref{fig:hallres}(b), where we measure the resistance over two lateral contacts. (b): The gray dots are the extracted upper critical field from the intersection of the gray line and the individual spline fits in (a). The black solid line is a fit to \cref{eq:Bc2T}, with $B_\text{c2}(0)$ and $T_\text{c0}$ as free parameters. The dashed line gives the slope at $T_\text{c0}$, $dB_\text{c2}/dT|_\text{T=Tc}=-2.0~\text{T/K}$.}
\end{figure*}
The only other parameter missing for the calculations of the scattering times and variance is the single spin density of states at the Fermi level, $N_0$. We use the Einstein relation, $N_0=1/(2e^2\rho_\text{N} D)$ to calculate $N_0$ from the diffusion constant $D$. This can be measured in superconductors from the temperature dependence of the upper critical field, $H_\text{c2}(T)$. Specifically \cite{Gershenzon1990, Kes1983}, 
\begin{equation}\label{eq:D}
	D = \frac{4k_\text{B}}{\pi e}\left(\left.\frac{d(\mu_0 H_\text{c2})}{dT}\right|_{T=T_\text{c0}}\right)^{-1}. 
\end{equation}
$T_\text{c0}$ is the critical temperature at zero field. To determine the slope of $B_\text{c2}(T)=\mu_0H_\text{c2}(T)$ at $T_\text{c0}$, we measure resistance versus temperature curves for different applied fields, as shown in \cref{fig:Hc2}(a). We use the same Hall bar geometry as shown in the inset of \cref{fig:hallres}(b) to measure the longitudinal resistance. We determine the critical temperature for a given magnetic field as the temperature where the resistance is 50\% of the normal state resistance, and invert that to get $B_\text{c2}(T)$. The result is shown in \cref{fig:Hc2}(b).\\
To account for a slight non-linearity of $B_\text{c2}(T)$ at $T_\text{c0}$, we fit $B_\text{c2}(T)$ to the function \cite{Tinkham2004},
\begin{equation}\label{eq:Bc2T}
	B_\text{c2}(T)=B_\text{c2}(0)\left(\frac{1-\left(T/T_\text{c0}\right)^2}{1+\left(T/T_\text{c0}\right)^2}\right),
\end{equation}
where $B_\text{c2}(0)$ is the upper critical flux density at $T=0$. We use both $B_\text{c2}(0)$ and $T_\text{c0}$ as fit parameters and we are interested in ratio, which gives the slope of $B_\text{c2}(T)$ at $T_\text{c0}$. The fit is shown in \cref{fig:Hc2}(b) as the solid black line together with the slope at $T_\text{c0}$ as dashed line, which is $dB_\text{c2}/dT|_{T=T_\text{c0}}=-1.5~\text{T/K}$.\\
This slope gives from \cref{eq:D} $D=0.74~cm^2/s$, and $N_0=2.1\times10^{4}~\mu\text{eV}^{-1}\mu\text{m}^{-3}$, which is comparable to the value $3.1\times10^{4}~\mu\text{eV}^{-1}\mu\text{m}^{-3}$ from Ref. \cite{Magnuson2019}. These values result in a Ginzburg-Landau coherence length of $\xi_\text{GL}=16~\text{nm}$.

\begin{figure*}[b]
	\includegraphics[width=.5\textwidth]{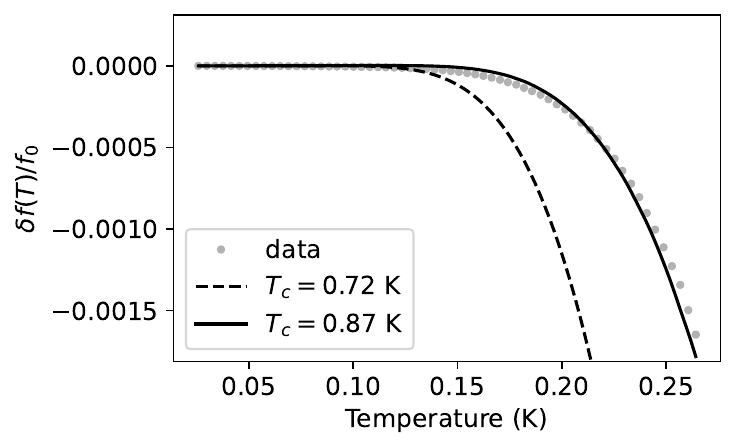}
	\caption{\label{fig:dff0} Fractional frequency shift with temperature of the membrane resonator compared to Mattis-Bardeen theory \cite{Mattis1958} to verify the critical temperature of $T_\text{c}=0.87~\text{K}$. The dashed and solid lines are calculated with the parameters in \cref{tab:params}, $\alpha_\text{k}^\text{sim}=0.66$ from simulation and $T_\text{c}$ as indicated by the legend.}
\end{figure*}

Surprisingly, we find from \cref{fig:Hc2}(b) $T_\text{c0}=0.72~\text{K}$, while we found $T_\text{c}=0.87~\text{K}$ in \cref{fig:hallres}(a). We verified this lower $T_\text{c}$ for the Hall bar structure in the setup without magnetic field which was used to obtain the data in the main text. The 4-probe and Hall bar structures are patterned in the same deposited film and close together on the wafer. The main difference is the width of the structures: 6 $\mu$m versus 200 $\mu$m, see the insets of \cref{fig:Hc2}(a) and (b). Why this would change the $T_\text{c}$ is not known, although it could be related to the stress in the film that might change the electron-phonon coupling for different widths. Nonetheless, we use the Hall bar geometry to extract normal state electronic properties only, namely the diffusion constant and the carrier charge density. For the diffusion constant, $D\propto T_\text{c}/B_\text{c2}$ with, in the dirty limit, $B_\text{c2}\propto 1/\xi_\text{GL}^2\propto 1/\xi_0 \propto T_\text{c}$, where $\xi_0$ the BCS coherence length. From this, we see that the diffusion constant does not depend on $T_\text{c}$ to first order. The carrier charge density is extracted in the normal state. Therefore, we expect that the different $T_\text{c}$ of the Hall bar does not have an effect on the extracted properties.\\
We take $T_\text{c}=0.87~\text{K}$ for the inductors, as they are $10~\mu\text{m}$ wide, which similar to the 4-probe structure width. We verify this $T_\text{c}$ by comparing the measured fractional resonance frequency shift versus temperature, $\delta f(T)/f_0$, to the calculated curves from Mattis-Bardeen \cite{Mattis1958} with $T_\text{c}=0.72~\text{K}$ and $T_\text{c}=0.87~\text{K}$ as input. We used a kinetic induction fraction of $\alpha_\text{k}^\text{sim}=0.66$ from a simulation, since the measurement of $\alpha_\text{k}=0.44$ assumes that $T_\text{c}=0.87~\text{K}$. This comparison is shown in \cref{fig:dff0}, from which we conclude that the inductors have a $T_\text{c}$ of $0.87~\text{K}$. The small deviations from the solid line are most likely due to the disorder in the film, as the complex sheet impedance of disordered superconductors is slightly different than predicted by standard Mattis-Bardeen theory \cite{Driessen2012,Khvalyuk2024}.

\section{T\lowercase{i}N data analysis}

\begin{figure*}[t]
	\includegraphics[width=.9\textwidth]{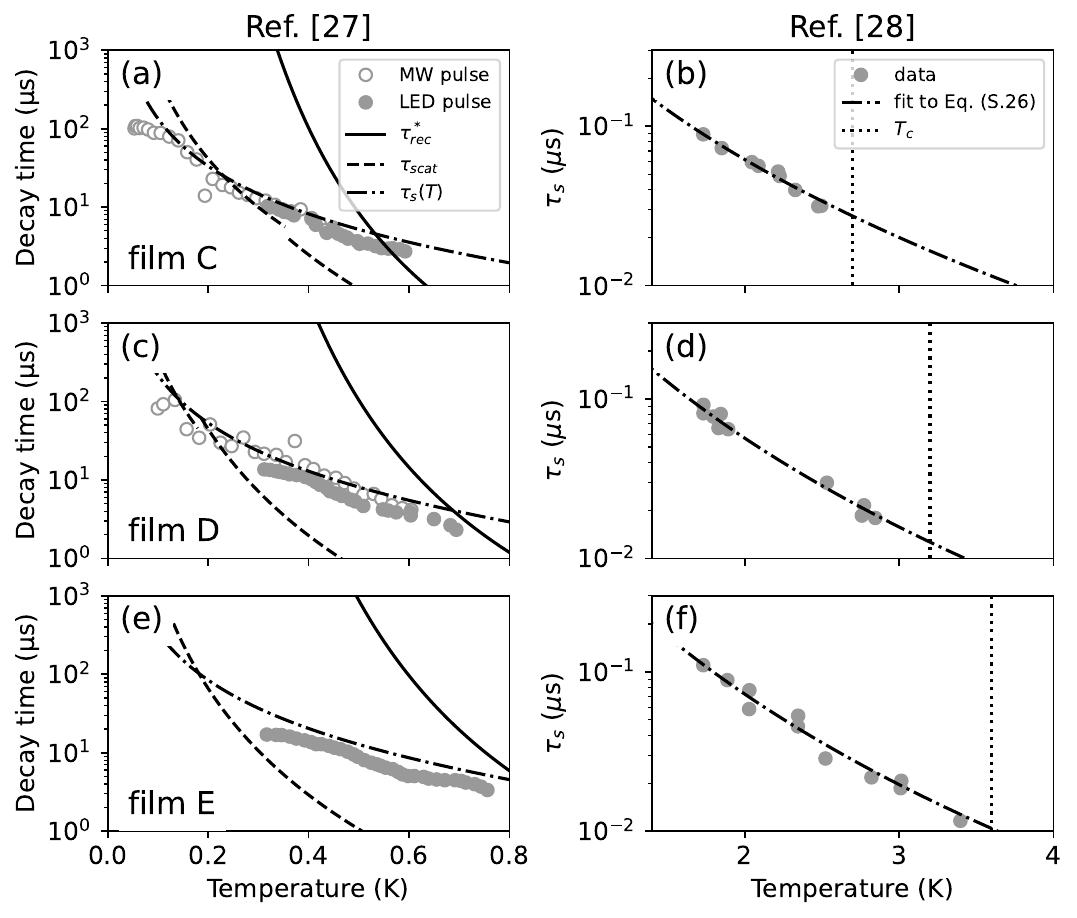}
	\caption{\label{fig:TiN} Relaxation times of three different TiN films from Refs. \protect\cite{Coumou2013a,Kardakova2015}, compared to disorder theory of Refs. \protect\cite{Devereaux1991a, Reizer1986, Sergeev2000}. (a), (c) and (e): Decay times of a pulse response from Ref. \protect\cite{Coumou2013a}. Solid gray dots are for a LED pulse and the open grey circles are for a microwave pulse applied at the resonator read-out line. The solid line is given by \cref{eq:taurecstar}, with $\tau_\text{rec}$ given by \cref{eq:trec_e-ph} and the phonon trapping factor given by \cref{eq:phtrf}. The dashed lines are from \cref{eq:tscat_e-ph2D} for low temperatures ($T<\hbar c_\text{T}/(2dk_\text{B})$) and from \cref{eq:tscat_e-ph3D} for high temperatures ($T>\hbar c_\text{T}/(2dk_\text{B})$).  (b), (d) and (f): The electron-phonon scattering time as measured in \protect\cite{Kardakova2015}, for the same films as (a), (c) and (e), respectively. The dotted vertical line indicates the critical temperature. The dashed-dotted black line is a fit to \cref{eq:tausT_meas}, with $k$ and $c_\text{L}$ as free parameters and assuming $c_\text{T}=c_\text{L}/2$. The dashed-dotted lines in (a), (b) and (c) are the same, but without correction factor (\cref{eq:tausT_meas}). The legends are applicable to all the figures in the same column. We used the parameters from \cref{tab:TiN} to calculate the theoretical lines.}
\end{figure*}

The electron-phonon scattering time has a weak power-law temperature dependence (\cref{eq:tauscat}), which contrasts the exponential temperature dependence of conventional quasiparticle recombination (\cref{eq:taurec}). This weak temperature dependence has also been measured with TiN resonators in Ref. \cite{Coumou2013a}. There, the relaxation time is measured as the decay time of a pulsed excitation, which is either generated by a (high photon number) LED pulse, or a microwave pulse applied to the read-out line of the resonators. The normal state electron-phonon times of these 3 films have been measured in Ref. \cite{Kardakova2015}, where they suppress $T_\text{c}$ using a magnetic field. The results of these measurements are shown in \cref{fig:TiN}. In this section, we analyze the three films from Ref. \cite{Coumou2013a}, C, D and E, in the same manner as we analysed the $\beta$-Ta film in the main text.

The parameters used for the calculations are shown in \cref{tab:TiN}. The first 6 columns are from Refs. \cite{Coumou2013a,Kardakova2015} and give us the electronic properties. From the $k_\text{F}l$ values we see that these films are indeed disordered, but somewhat less than the $\beta$-Ta film from the main text. To estimate the rest of the electronic properties, we need $N_0$ as well. This has been measured in Ref. \cite{Kardakova2013} for similar films. We compare the thickness of each film and estimate $N_0$ to be the values in \cref{tab:TiN}. With these values, we calculate $\xi/d\lesssim0.25$. So, these are 3D superconducting films.

We estimate the phonon properties of the TiN films in the following way. The mass density we get from Ref. \cite{Hansen2020}, for TiN film with similar thickness. The Debye temperature we get from Ref. \cite{Chen2009}. The transverse and longitudinal sound velocities, we estimate from a fit to the data for Ref. \cite{Kardakova2015}. This is procedure is also set out in Ref. \cite{Sidorova2020}. We assume $c_\text{T}=c_\text{L}/2$ and let $c_\text{L}$ be a free fit parameter. This relation seems to be correct for many materials \cite{Kaplan1976}. The results do not change if we assume $c_\text{T}<c_\text{L}/2$, since the relaxation times are then limited by transverse phonon scattering and we are left with only one sound velocity. To fit the measured electron-phonon relaxation times, we use the theory of Ref. \cite{Sergeev2000}, which includes statics scatterers when $k\leq 1$. We use Eq. (51) of \cite{Sergeev2000},
\begin{align}\label{eq:tausT}
	\begin{split}
	\frac{1}{\tau_\text{s}(T)}=&\frac{\pi^4\beta_\text{L}}{5\hbar^4}\frac{(k_\text{F} l) (k_\text{B} T)^4 }{(k_\text{F} c_\text{L})^3}\left(1+k\frac{3}{2}\left(\frac{c_\text{L}}{c_\text{T}}\right)^5\right) \\
	&+ \frac{3\pi^3\beta_\text{L}}{2\hbar^2}\frac{(k_\text{B}T)^2}{(k_\text{F}l)(k_\text{F} c_\text{L})}(1-k)\left(1 + 2k\left(\frac{c_\text{L}}{c_\text{T}}\right)^3\right).
	\end{split}
\end{align}
$\beta_\text{L}$ is defined in \cref{eq:betaL}. These measurements are performed at a finite temperature, above $T_\text{c}$, where the electrons distributed in energy around the Fermi surface with a width of $k_\text{B}T$. The measured signal is an average of all these electrons. So, the measured $\tau_\text{s}^\text{meas.}$, is given by \cite{Sidorova2020,Ilin1998},
\begin{equation}\label{eq:tausT_meas}
	\frac{1}{\tau_\text{s}^\text{meas.}}(T) = \frac{3(n+2)\Gamma(n+2)\zeta(n+2)}{2\pi^2(2-2^{1-n})\Gamma(n)\zeta(n)}\frac{1}{\tau_\text{s}(T)}. 
\end{equation}
For $n$ we use the values from \cite{Kardakova2015}, see \cref{tab:TiN}. We use \cref{eq:tausT_meas} to fit the the data of \cite{Kardakova2015}, as shown in \cref{fig:TiN}(b), (d) and (f). The fit results are given in the two right columns of \cref{tab:TiN}.

\begin{table*}[t]
	\caption{\label{tab:TiN} Parameters for the TiN films from Refs. \protect\cite{Coumou2013a,Kardakova2015}, with additional parameters estimated from Refs. \cite{Kardakova2013,Hansen2020,Chen2009}. $N_0$ and $\hat{\rho}$ are estimated from similar films taking the thickness into account. $n$ is the power of the temperature dependence of $\tau_\text{s}$, as measured in \protect\cite{Kardakova2015}. $c_\text{L}$ and $k$ are the results of the fits in \cref{fig:TiN}(b), (d) and (f), using \cref{eq:tausT_meas} and assuming $c_\text{T}=c_\text{L}/2$.}
	\begin{ruledtabular}
		\begin{tabular}{c|ccccc|c|c|c|ccc}
			\multicolumn{5}{c}{From Refs. \cite{Coumou2013a,Kardakova2015}}&             & Ref. \cite{Kardakova2013} &           Ref. \cite{Hansen2020}            &       Ref. \cite{Chen2009}       &            \multicolumn{2}{c}{Fit Ref. \cite{Kardakova2015} to \cref{eq:tausT_meas}} &  \\
			Film & $d$ (nm) & $T_\text{c}$ (K) & $\rho_\text{N}$ ($\mu\Omega$cm) & $k_\text{F} l$ & $n$                       & $N_0$ ($\mu\text{eV}^{-1}\mu\text{m}^{-3}$) & $\hat{\rho}$ (g/cm\textsuperscript{3}) & $T_\text{D}$ (K) & $c_\text{L}$ (km/s) &                             k                              &  \\ \hline
			C   &    22    &    2.7    &           253            &   4.6   & 2.8                       &              $6.2\times10^{4}$              &               5.7                &    579    &     4.0      &                           0.993                            &  \\
			D   &    45    &    3.2    &           187            &   6.1   & 3.1                       &              $6.0\times10^{4}$              &               5.0                &    579    &     4.2      &                           0.995                            &  \\
			E   &    89    &    3.6    &           120            &   8.6   & 3.3                       &              $5.9\times10^{4}$              &               5.0                &    579    &     5.1      &                           0.995                            &
		\end{tabular}
	\end{ruledtabular}
\end{table*}

With also the sound velocities available, we can estimate the phonon dimensionality via $qd$ and disorder via $ql$. For thermal phonons, we find that $q(k_\text{B}T)d=1/2$ at 0.36 K, 0.18 K and 0.11 K, for film C, D and E respectively. Below these temperatures, the thermal phonons are 2D, above them they are 3D. Recombination phonons are 3D, $q(2\Delta_0)d\geq 13$ for all films.\\ 
Electron phonon interaction is disordered, since both $q(2\Delta_0)l\leq0.46$ and $q(0.8~\text{K})l\leq0.03$ are smaller than one. We therefore use the same equations as for $\beta$-Ta, namely \cref{eq:trec_e-ph} for recombination and \cref{eq:tscat_e-ph2D,eq:tscat_e-ph3D} for scattering.\\
We calculate the phonon trapping factor in the same manner as for $\beta$-Ta, i.e. \cref{eq:phtrf}, but with the parameters in \cref{tab:TiN}. The substrate is in this case Si, for which we use: $\hat{\rho}=2.33~\text{g/cm}^3$, $c_\text{L}=8.98~\text{km/s}$ and $c_\text{T}=5.34~\text{km/s}$ \cite{Kaplan1979}. The diffuse scattering assumption is in this case justified by the presence of a native oxide layer between the TiN and the Si. Furthermore, the TiN films are poly-crystalline with a grain size smaller than 42 nm \cite{Coumou2013a}. We get from \cref{eq:phtrf} a phonon trapping factor of 29, 96 and 193 for films C, D and E, respectively. As an example, for film D the pair breaking time is on the order of 2 ns and phonon escape time on the order of 200 ns.

The resulting scattering and recombination times are shown in \cref{fig:TiN}(a), (c) and (e) by the dashed and solid lines. The measured quasiparticle relaxation times have an even weaker temperature dependence than the phonon scattering time, $n<9/2$. However, the theory of Refs. \cite{Reizer1986,Devereaux1991a} do not take static scattering into account, while we see from the fit results in \cref{tab:TiN} that $k<1$ for all films. A theory for inelastic electron-phonon scattering in superconductors that includes static scatterers is still lacking. Nonetheless, we can compare the quasiparticle relaxation times to the normal state scattering time, $\tau_\text{s}(T)$. From \cref{eq:tausT} we see that static scatters cause a $n=2$ power-law when $k<1-(k_\text{B}T l/ (\hbar c_\text{T}))^2<0.999$ \cite{Sergeev2000}, which is satisfied in all experimental regimes in \cref{fig:TiN}(a), (c) and (e). We plotted $\tau_\text{s}(T)$ from \cref{eq:tausT} with the same film parameters and no fit parameter. We omitted the averaging factor in \cref{eq:tausT_meas}, as the quasiparticles in a superconductor all have an energy very close to $\Delta_0$ due to the gap in the density of states.\\
We observe that the relaxation times follow the same $n=2$ temperature dependence for all film. The disorder dependence of the relaxation rate is also captured by \cref{eq:tausT}, since the difference between the data and \cref{eq:tausT} is constant for the variation in disorder in the different films. We therefore conclude that also in TiN, the quasiparticle relaxation is enhanced and governed by the phonon scattering time. It also shows that this phenomenology is not limited to the $\beta$-Ta film we study in the main text, but is present in different disordered superconductors.

If we take $\xi^\text{TiN}\approx6~\text{nm}$ as localization length, we get a density of localized quasiparticle states of $n_\text{qp}^\text{loc}=3/(4\pi (\xi^\text{TiN})^3)\approx 1\times10^6~\mu\text{m}^{-3}$. The thermal quasiparticle density at the maximum measured temperatures in \cref{fig:TiN}(a), (c) and (e) is $n_\text{qp}^T\approx 7 \times 10^3~\mu\text{m}^{-3}$. This shows we are in the same regime as the $\beta$-Ta film of the main text: the thermally excited quasiparticles localize and recombine after delocalization by electron-phonon scattering. The values mentioned are based on the parameters in \cref{tab:TiN} and are the same within a factor 2 for all three films.

\section{Fitting the power spectral densities}\label{sec:fit}
\begin{figure*}[t]
	\includegraphics[width=.9\textwidth]{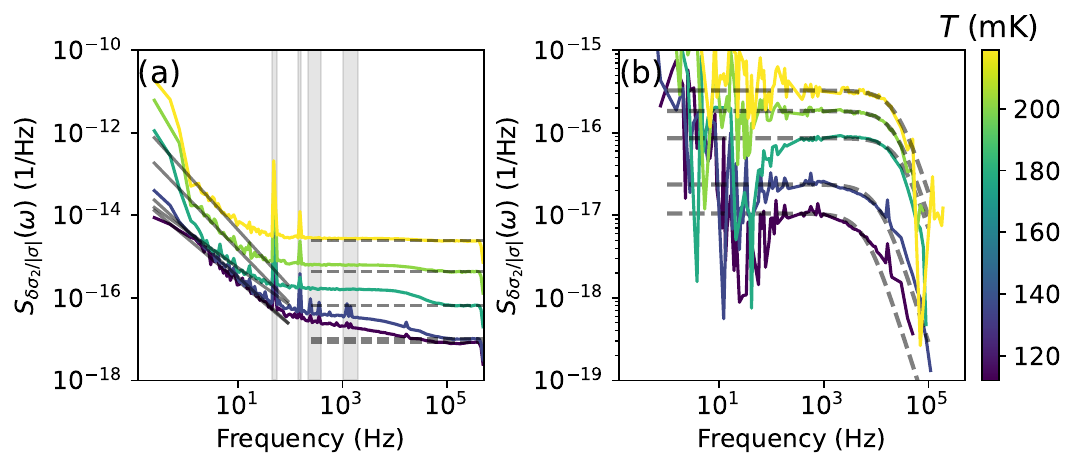}
	\caption{\label{fig:specfilter} Example of the subtraction of other noise sources and fitting of the measured kinetic inductance fluctuation power spectral density (PSD). (a): Measured raw PSD as solid lines for the substrate resonator at different temperatures, indicated by the colors. The gray shaded vertical areas indicate the frequency ranges that we disregard to limit the influence of 50 Hz spikes. The dashed gray lines give the maxima of the solid lines between $250~\text{kHz}$ and $500~\text{kHz}$, which we subtract as amplifier noise. The solid gray lines give a $A/f^\gamma$-fit, with $A$ and $\gamma$ as fit parameters, which we subtract to limit the effects of $1/f$ noise. (b): Resulting PSDs when the other noise contributions are subtracted. The dashed gray lines give the Lorentzian fit, \cref{eq:Sfit}, from which we extract the quasiparticle lifetime and variance (\cref{fig:results}(a) and (b)).}
\end{figure*}

\Cref{fig:specfilter}(a) shows 5 examples of measured power spectral densities (PSDs) for the substrate resonator. Contributions from 50 Hz interference, $1/f$ fluctuators and the 4 K HEMT amplifier are clearly visible in the PSDs. To minimize the effects from these noise sources to the Lorentzian fit, we first disregard the data points in the gray shaded regions in \cref{fig:specfilter})(a). Second, we take the maximum value between $250~\text{kHz}$ and $500~\text{kHz}$ as the amplifier noise contribution and subtract that from the PSD. This value is shown by the grey dashed line. Third, we fit the function, $A/f^\gamma$, from $3~\text{Hz}$ to $100~\text{Hz}$ with $A$ and $\gamma$ as fit parameters. The fit is shown as the gray solid line, which is also subtracted from the PSD. The resulting PSDs are shown in \cref{fig:specfilter}(b). We fit \cref{eq:Sfit} to these PSDs, which results in the gray dashed lines and gives us $\tau$ and $s^2$ in \cref{fig:results}(a) and (b).

\section{Fluctuation model for localized quasiparticle recombination}
\begin{figure}
	\includegraphics[width=\textwidth]{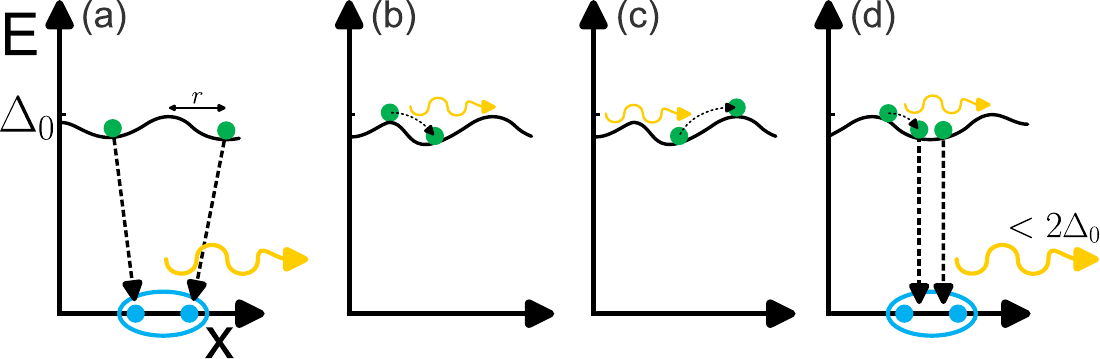}
	\caption{\label{fig:locrecsketch} Sketches of the different transitions in the rate equations, \cref{eq:rateeqs}. (a): Localized quasiparticle recombination, as described in Ref. \cite{Bespalov2016a}. The recombination rate depends on the distance between two quasiparticles, $2r$. The emitted phonon has less than $2\Delta_0$ energy at low temperatures, but can be $\geq2\Delta_0$ for higher temperatures when the quasiparticles are not fully relaxed into the localized states. (b): Quasiparticle localization with emission of a phonon. (c): Quasiparticle delocalization via absorption of a phonon. (d): On-site recombination, where a mobile quasiparticle relaxes to a localized quasiparticle in the same localization site. This process emits multiple phonons with energy $<2\Delta_0$. Phonons are depicted with curvy yellow arrows, quasiparticles with green dots and Cooper-pairs with blue, grouped dots.}
\end{figure}

\begin{figure}
	\includegraphics[width=\textwidth]{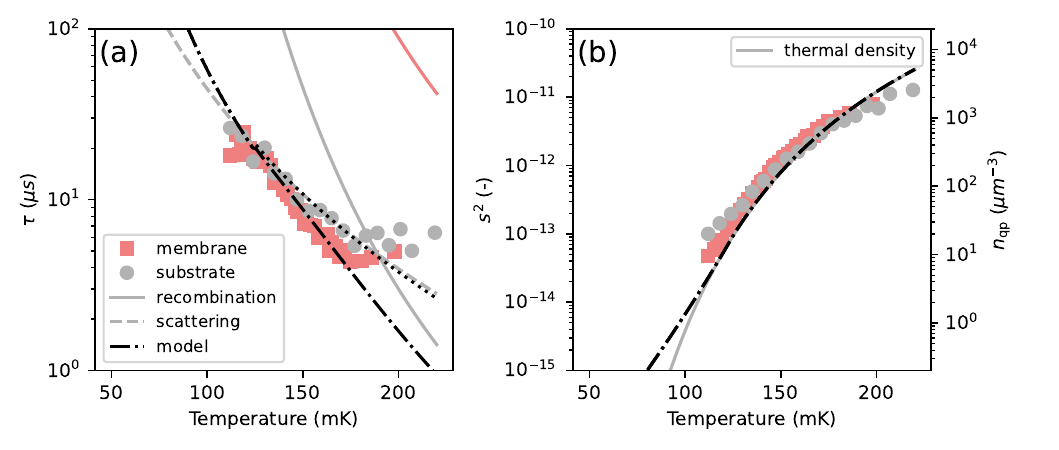}
	\caption{\label{fig:tauvar} Calculated relaxation time and variance from the fluctuation model, compared to the measured data. (a): The solid lines give the localized recombination time ($1/(4R_\text{l}n_\text{l}^0)$), which corresponds to conventional recombination in this regime. The gray solid line is for the substrate resonator and the red line is for the membrane resonator. The gray dashed line is the delocalization time ($\tau_\text{scat}/2$). The solid and dashed lines are identical to the lines in \cref{fig:results}(a) in the main text. The black dashed-dotted lines are the model prediction for the substrate resonator. This is given by the reciprocal of the smallest eigenvalue of $\underline{\underline{M}}$. The dotted black line is for the membrane case. This follows $\tau_\text{scat}/2$ to much higher temperatures than in the substrate case, since the localized recombination is much slower (red solid line). (b): The solid gray line corresponds to the thermal quasiparticle density (\cref{eq:nqpT}). The dashed-dotted line is calculated from the model as the sum of the elements of  $\underline{\underline{s^2}}=\underline{\underline{M}}^{-1}\underline{\underline{B}}/2$. This is equal to $n_\text{qp}/V=C_{p}/(4\pi V/3r^3)$, with $r$ from \cref{eq:r}. For temperatures $T>124~\text{mK}$, this equals $n_\text{qp}^T/V$ from \cref{eq:nqpT}. For lower temperatures, the deviation of the dashed-dotted line from the thermal density shows that quasiparticle localization effects cause an excess quasiparticle density as predicted in Ref. \cite{Bespalov2016a}. The gray dots and red squares in both panels are the same data points as in \cref{fig:results} in the main text.}
\end{figure}

From \cref{fig:results} in the main text, we observe that the measured complex conductivity fluctuations can be interpreted as quasiparticle number fluctuations \cite{deVisser2011}, but with a lifetime of $\tau_\text{scat}/2$, instead of the conventional $\tau_\text{rec}^*$ (\cref{eq:taurecstar}). We conclude that the recombination of two quasiparticles must be preceded with a phonon absorption event. We explain this by recombination of localized quasiparticles, that first delocalize via phonon absorption, see \cref{fig:locrecsketch}(c) and (d). In this section, we model the quasiparticle fluctuations with the master equation approach as described in Refs. \cite{Wilson2004,VanVliet1965}, including the localized quasiparticle dynamics described in Ref. \cite{Bespalov2016a}. At the end of this section, we will discuss the assumptions and implications of the model.\\
To capture the observed phenomenology, we use the following rate equations for mobile and localized quasiparticle densities,
\begin{equation}\label{eq:rateeqs}
	\begin{alignedat}{2}
	\frac{dn_\text{m}}{dt} &= &-&\Gamma_\text{l} n_m + \Gamma_\text{d} n_l -2 R_\text{os} \left(n_\text{m}n_\text{l} - n_\text{m}^0n_\text{l}^0\right)\\
		\frac{dn_\text{l}}{dt} &= - 2 R_\text{l} n_\text{l}^2 + 2\Gamma_\text{pb} &+& \Gamma_\text{l} n_m -\Gamma_\text{d} n_l.
	\end{alignedat}
\end{equation}
Here, $n_\text{m}$ is the mobile quasiparticle density and $n_\text{l}$ is the localized quasiparticle density. The first two terms in the rate equation for $n_\text{l}$ (second line) are the \textit{localized} quasiparticle recombination (\cref{fig:locrecsketch}(a)) and generation terms described in Ref. \cite{Bespalov2016a}. We assume that the localized recombination and pair-breaking are in detailed balance ($2R_\text{l}n_\text{l}^2=2\Gamma_\text{pb}$), as we do for all other transitions, since we are in equilibrium \cite{Wilson2004}. The main evidence for this is that the fluctuation variance equals the variance of a thermal quasiparticle density (\cref{fig:results}(b)).\\
We set the generation rate equal to the pair-breaking rate of thermal phonons, $\Gamma_\text{pb}=R \left(n_\text{qp}^T\right)^2/2$, since we measure at a relatively high bath temperatures. $R$ is the conventional recombination constant, $R=\left(2\Delta_0/(k_\text{B}T_\text{c})\right)^3/(4 N_0\Delta \tau_0^\text{rec})$, with $\tau_0^\text{rec}=21~\text{ns}$ for $\beta$-Ta \cite{Reizer1986,Kaplan1976} (see Supplementary Section 1). The thermal quasiparticle density is given by,
\begin{equation}\label{eq:nqpT}
	n_\text{qp}^T=2 N_0 \Delta_0 \sqrt{\frac{2 \pi k_\text{B}T}{\Delta_0}} e^{-\Delta_0 / k_\text{B}T}.
\end{equation} \\
The recombination rate for localized quasiparticles (\cref{fig:locrecsketch}(a)) depends on the average distance between quasiparticles, $2r$ \cite{Bespalov2016a}, 
\begin{equation}\label{eq:Rl}
	R_\text{l} = 
	\begin{cases}
		\dfrac{R}{2}\left(\dfrac{4\pi}{3 C_\text{p}}\right)^2 b \left(\dfrac{r}{r_\text{c}}\right)^{\beta+3} e^{-r/r_\text{c}} & r/r_\text{c} > 3\\[18pt]
		\dfrac{R}{2(1+\tau_\text{esc}/\tau_\text{pb})} & r/r_\text{c} < 3
	\end{cases}.
\end{equation}
Here, $b=0.008$ and $\beta=0.41$ \cite{Bespalov2016a}. In Supplementary Section 6, we estimate the critical quasiparticle distance to be $r_\text{c}\simeq 3.5\xi=65~\text{nm}$. We obtain $r$ from numerically solving the equation \cite{Bespalov2016a}, 
\begin{equation}\label{eq:r}
	b\left(\frac{r}{r_\text{c}}\right)^{\beta-3}e^{-r/r_\text{c}} = \frac{\Gamma_\text{pb}r_\text{c}^6}{R}.
\end{equation}
Since $\Gamma_\text{pb}$ is temperature dependent, this results in a temperature dependent quasiparticle density, $n_\text{qp}=C_{p}/(4\pi/3r^3)$. For $r/r_\text{c}\lesssim 3.0$, the quasiparticle density should be equal to the thermal number of quasiparticles, $n_\text{qp}(r/r_\text{c}<3)\approx n_\text{qp}^T$ \cite{Bespalov2016a}. Therefore, we set $R_\text{l}=R/(1+\tau_\text{esc}/\tau_\text{pb})$ when $r/r_\text{c}<3$ in \cref{eq:Rl}. We include the effect of phonon trapping for $r<3r_\text{c}$ and omit it for $r>3r_\text{c}$, since the quasiparticles are likely to have an energy close to $\Delta_0$ only at higher temperatures. Furthermore, we set the quasiparticle packing coefficient $C_p=0.54$, such that $n_\text{qp}(r/r_\text{c}=3)=n_\text{qp}^T$. This is slightly smaller than the value $C_p=0.61$ mentioned in Ref. \cite{Bespalov2016a}. With these parameters, we obtain a cross-over temperature for $r/r_\text{c}=3$ of $T\approx124~\text{mK}$. For lower temperatures we expect an excess number of quasiparticles due to localization. For higher bath temperatures, we expect a thermal quasiparticle density, which is consistent with the measured variance in \cref{fig:results}(b) in the main text (see also \cref{fig:tauvar}(b)). 

If we would only include the localized quasiparticle recombination ($R_\text{l}$) and thermal generation ($\Gamma_\text{pb}$) terms, as is considered in Ref. \cite{Bespalov2016a}, we would obtain the conventional quasiparticle fluctuations in our measurement regime, with a thermal quasiparticle density as variance and $\tau_\text{rec}^*$ as relaxation time. We expect however that quasiparticles can delocalize due to inelastic phonon scattering at these relatively high bath temperatures. This enables a second, faster recombination mechanism: \textit{on-site} recombination (\cref{fig:locrecsketch}(d)). To include this in the model, we introduce a mobile quasiparticle level and localization and delocalization terms in \cref{eq:rateeqs}, see \cref{fig:locrecsketch}(b) and (c). We set the delocalization rate to the reciprocal of the inelastic phonon scattering time, $\Gamma_\text{d}=1/\tau_\text{scat}$, which is the time it takes to absorb a thermal phonon (\cref{fig:locrecsketch}(c)). Since we assume that we are in equilibrium, localization and delocalization obey detailed balance and the steady state values are given by, 
\begin{equation}\label{eq:ss}
	\begin{aligned}
		n_\text{l}^0 &= n_\text{qp} \\
		n_\text{m}^0 &= n_\text{qp}\frac{\Gamma_\text{d}}{\Gamma_\text{l}},
	\end{aligned}
\end{equation}
We set the localization rate (\cref{fig:locrecsketch}(b)) much higher than the delocalization rate, $\Gamma_\text{d}\ll\Gamma_\text{l}=10^2/\tau_\text{scat}$, such that the mobile quasiparticle density is low and the total density is close to $n_\text{qp}$. We omit the conventional mobile-mobile recombination, since these rates are very low at a low mobile quasiparticle density.\\
The last term in \cref{eq:rateeqs}, first line, describes the relaxation of a mobile quasiparticle and subsequent \textit{on-site} recombination with a localized quasiparticle (\cref{fig:locrecsketch}(d)). The rate depends on both the mobile and localized quasiparticle density, $n_\text{m}$ and $n_\text{l}$, and we include the $n_\text{m}^0n_\text{l}^0$ term to ensure that detailed balance is satisfied. The on-site recombination constant is given by $R_\text{os}=\left(2\Delta/(k_\text{B}T_\text{c})\right)^3/(4 N_0\Delta \tau_0^\text{os})$, with $\tau_0^\text{os}$ the characteristic on-site recombination time. This is likely to be short, since the mobile and localized quasiparticles relax to the same location \cite{Bespalov2016a}. We choose this characteristic time to be $\tau_0^\text{os}=10^{-4}\tau_0^\text{rec}$, such that on-site recombination is faster than localization and delocalization over our entire measurement regime. Any shorter characteristic time gives the same results when comparing it to our measurements. We disregard the phonon trapping effect for on-site recombination (\cref{fig:locrecsketch}(d)), since this is a multi-step relaxation process and the individual phonons emitted have less energy than $2\Delta_0$ and are unlikely to break a Cooper-pair \cite{Kozorezov2008}.\\

To obtain the quasiparticle fluctuation spectra, we calculate the two matrices $\underline{\underline{M}}$ and $\underline{\underline{B}}$, that describe the drift and diffusion in the $(n_\text{m}, n_\text{l})$-space, respectively \cite{Wilson2004}. From \cref{eq:rateeqs}, we read off the transition probabilities, $p$, and shot sizes, $\delta n$,
\begin{equation}\label{eq:transprob}
	\begin{alignedat}{3}
		p_{12}&=\Gamma_\text{l}n_\text{m} &\qquad &\delta n_{12} = 1/V \\
		p_{13}&=R_\text{os} n_\text{m} n_\text{l}&\qquad &\delta n_{13} = 2/V\\
		p_{21}&=\Gamma_\text{d}n_\text{l} &\qquad &\delta n_{21} = 1/V \\
		p_{23}&=R_\text{l}n_\text{l}^2 &\qquad &\delta n_{23} = 2/V \\
		p_{31}&=R_\text{os} n_\text{m}^0 n_\text{l}^0 &\qquad &\delta n_{31} = 2/V\\
		p_{32}&=\Gamma_\text{pb} &\qquad &\delta n_{31} = 2/V, \\
	\end{alignedat}
\end{equation}
where the index 1 denotes the mobile quasiparticle level, 2 the localized quasiparticle level and 3 the Cooper-pair level. From this, we can directly calculate the matrices with \cite{Wilson2004}
\begin{equation}
	\begin{alignedat}{1}
		M_{ij} &= \sum_k \delta n_{ik} \left( \left. \frac{\partial p_{ik}}{\partial n_j} - \frac{\partial p_{ki}}{\partial n_j} \right) \right|_{\{n_i\} = \{n_i^0\}}\\
		B_{ii} &= \sum_{k \neq i} \delta n_{ik}^2 (p_{ki}^0 + p_{ik}^0) \\
		B_{ij} &= - \delta n_{ij} \delta n_{ji} (p_{ij}^0 + p_{ji}^0).
	\end{alignedat}
\end{equation}
Using these equations results in,
\begin{equation}
	\begin{alignedat}{1}
		\underline{\underline{M}} &= \frac{1}{V}
		\begin{bmatrix}
			\Gamma_\text{l} + 2 R_\text{os} n_\text{l}^0 & -\Gamma_\text{d} + 2 R_\text{os} n_\text{m}^0 \\
			-\Gamma_\text{l} & \Gamma_\text{d} +4R_\text{l}n_\text{l}^0
		\end{bmatrix}\\
		\underline{\underline{B}} &= \frac{2 n_\text{l}^0}{V^2}
		\begin{bmatrix}
			 \Gamma_\text{d} + 4R_\text{os} n_\text{m}^0 & -\Gamma_\text{d}\\
			-\Gamma_\text{d} &  \Gamma_\text{d} + 4R_\text{l}n_\text{l}^0
		\end{bmatrix},
	\end{alignedat}
\end{equation}
where we used the principle of detailed balance (i.e. $p_{ij}^0 = p_{ji}^0$) in the expression for $\underline{\underline{B}}$. 

The eigenvalues of the matrix $\underline{\underline{M}}$ give the fluctuation rates. In the limit of slow localized recombination and fast on-site recombination, $R_\text{l}n_\text{l}^0\ll\Gamma_\text{d}\ll\Gamma_\text{l}\ll R_\text{os}n_\text{m}^0$, the smallest fluctuation rate is $1/\tau=2\Gamma_\text{d}=2/\tau_\text{scat}$. This corresponds to the temperature regime $T<175~\text{mK}$ in \cref{fig:results} from the main text, where recombination is slow (black line) and we observe $2/\tau_\text{scat}$ (dashed line) as relaxation rate. The factor two in the relaxation rate reflects the pair-wise nature of recombination.\\
The variance matrix in equilibrium can be calculated as $\underline{\underline{s^2}}=\underline{\underline{M}}^{-1}\underline{\underline{B}}/2$ \cite{Wilson2004}. We compare the sum of the matrix entries to the experimental measured variance ($s^2=\sum_{i,j}s^2_{ij}$) since we are sensitive to both localized and mobile quasiparticles. This results in $s^2=(n_\text{l}^0 + n_\text{m}^0)/V \approx n_\text{qp}/V$, which is equal to $n_\text{qp}^T/V$ in our measurement regime. To summarize this analytical argument, we expect quasiparticle fluctuations with a relaxation time of $\tau=\tau_\text{scat}/2$ and a variance $s^2=n_\text{qp}^T/V$, which we observe from the experiment in \cref{fig:results} in the main text.\\ \Cref{fig:tauvar} shows the numerically calculated relaxation time and variance from this model. For this, we used the parameters as previously mentioned: $\Gamma_\text{l}=10^2\Gamma_\text{d}=10^2/\tau_\text{scat}$, $\tau_0^\text{os}=10^{-4}\tau_0^\text{rec}=10^{-4}\cdot21~\text{ns}$ and $R_\text{l}$ from \cref{eq:Rl}. This describes the data accurately, except for temperatures above $175~\text{mK}$. In that regime, the average distance between quasiparticles is larger than the critical distance, $r>r_\text{c}$. Therefore, a significant fraction of quasiparticles will be mobile, which is not included in our model. That suggests that the plateau in relaxation time we observe for $T>175~\text{mK}$ might be a cross-over regime from localized to mobile quasiparticle dynamics.

The full power spectral density matrix can be calculated with \cite{Wilson2004}, 
\begin{equation}\label{eq:PSDs}
	\underline{\underline{G}}(\omega) = \frac{2}{\omega^2} \operatorname{Re} \left[ \left( \mathbf{1} + {\underline{\underline{M}}}/{(i\omega)} \right)^{-1} \underline{\underline{B}} \right].	
\end{equation}
Here, $\omega$ is the angular frequency of the fluctuations and $\mathbf{1}$ is the identity matrix. Taking the sum of all matrix elements ($G(\omega) = \sum_{i,j}G_{ij}(\omega)$) results in the power spectral densities as shown in \cref{fig:specs}(c) and (d). These resemble the data accurately and also show the features $\tau=\tau_\text{scat}/2$ and a variance $s^2=n_\text{qp}^T/V$.\\

\subsection*{Discussion of assumptions and implications of the model}
At this point, we like to make some remarks about the assumptions in the model. First of all, we assumed that the delocalization time is given by the inelastic phonon scattering time ($\tau_\text{scat}$ from \cref{eq:tscat_e-ph2D}). This scattering time \cite{Kaplan1976,Reizer1986} is for quasiparticles with energy $\Delta_0$ and a BCS density of states. Since these quasiparticles cannot loose energy (the density of states below $\Delta_0$ energy in a BCS superconductor is 0), the only contribution to this inelastic interaction time is phonon absorption. This justifies the choice $\Gamma_\text{d}=1/\tau_\text{scat}$. 

Secondly, for the same reason as mentioned above, the scattering rate for phonon emission is zero for quasiparticles in BCS superconductor with $\Delta_0$ energy. For disordered superconductors, with a slightly altered density of states, it is thus likely that phonon emission by a mobile quasiparticle is very slow. We assumed however a much higher localization rate than delocalization, since that is required for most quasiparticles to be localized. It is not clear what causes this fast relaxation. Electron-electron interaction could be responsible for this, since it is enhanced in disordered metals and superconductors (see \cref{fig:lifetimes}). For a proper analysis however, these interaction times should be calculated with the full density of states that is affected by disorder.

Thirdly, since the density of states is much larger close to $\Delta_0$ (see \cref{fig:disrecsketch}(d)), thermal phonons are more likely to generate mobile quasiparticles. We however included this term ($\Gamma_\text{pb}$) in the localized quasiparticle level, to ensures detailed balance with the localized recombination ($R_\text{l}$). The conventional (mobile) quasiparticle dynamics coincides with these localized recombination and generation terms, since the localized quasiparticle density is high ($r<3r_\text{c}$, see \cref{eq:Rl} and Ref. \cite{Bespalov2016a}). Therefore, the conventional generation-recombination quasiparticle dynamics are also captured in this model, when fast localization is assumed (see second discussion point above). 

Fourthly, we modeled the on-site recombination as a rate that only affects the mobile quasiparticle density, while it would be more intuitive to include it in both the localized and mobile rate equations (see \cref{fig:locrecsketch}(d)). In fact, we could divide the $R_\text{os}$ term in \cref{eq:rateeqs} by 2 and add the same term to the $n_\text{l}$-rate equation. This results in a relaxation time of $\tau_\text{scat}/4$ and a variance of $n_\text{qp}/(2V)$, i.e., both missing a factor 2 compared to the measurements. The reason for that is that the pair-wise nature of on-site recombination is not captured in the model that way. It is not clear how to include such a multi-level, pair-wise transition in the framework of Ref. \cite{Wilson2004}. That is why we choose to include the on-site recombination term in the $n_\text{m}$-level only. \\
In a more rigorous treatment of the localization and recombination dynamics (\cref{fig:locrecsketch}), the effects of the local variations should be included. For example, position-dependent quasiparticle densities, $n_\text{m}(\vec{r})$ and $n_\text{l}(\vec{r})$, and a diffusion term in the mobile quasiparticle rate equation could be added. In that way, the localized and on-site recombination terms (\cref{fig:locrecsketch}(a) and (d)) could be combined in one recombination term that is dependent on the local quasiparticle density. This could give a microscopic reasoning for the fast on-site recombination characteristic time, which we here have set to $\tau_0^\text{os}=10^{-4}\tau_0^\text{rec}$. Such position-dependent models are much more complicated to solve, and we therefore do not treat them here. However, similar models have been developed in the context of fluctuations in semiconductors \cite{VanVliet1965,VanVliet1956}. In semiconductors, Shockley-Read-Hall trap-assisted recombination \cite{Shockley1952, Hall1952} of electrons and holes cause a similar enhanced recombination rate.

Lastly, we would like to comment on the low-temperature behavior of the model. For the quasiparticle density, the model follows the predictions of Ref. \cite{Bespalov2016a}: when the distance between the localized quasiparticles becomes larger than $6r_\text{c}$ localized recombination ($R_\text{l}$) slows down and causes an excess quasiparticle density (\cref{fig:tauvar}(b)). At some point, the pair-breaking rate of thermal phonons will be smaller than another pair-breaking mechanism such as cosmic rays, radioactivity or stray light. For our measurement, it is likely that the microwave read-out power is the dominating pair-breaking processes at low temperatures \cite{deVisser2012a}, see also \cref{fig:Pread}. At low temperatures, the quasiparticle density will therefore saturate completely.\\
Additionally, the on-site recombination slows down at low temperatures as $n_\text{qp}$ decreases. Therefore, a delocalized quasiparticle will be more likely to localize again, instead of recombine on-site. In the limit $\Gamma_\text{l}\ll R_\text{os}n_\text{l}^0$, relaxation is dominated by localized recombination (\cref{fig:locrecsketch}(a)) instead of on-site recombination (\cref{fig:locrecsketch}(d)). Localized recombination is more-than exponentially slow at low quasiparticle densities, with a relaxation time $\tau=1/(4R_\text{l}n_\text{l}^0)$ \cite{Bespalov2016a}. This could explain the measurements of Ref. \cite{Grunhaupt2018}, where relaxation times on the order of seconds have been measured in disordered grAl.\\ 

\section{Gap fluctuations and localized states induced by disorder}
To estimate the density of localized quasiparticle states, we need an estimate for $\eta$ and $\Gamma_\text{tail}$, which parameterize the disorder-induced broadening and exponential subgap tail of the density of states (see \cref{fig:disrecsketch}). Since $\xi/d=0.39$ (see \textit{Film characterization} in the \textit{Methods section}) is close to 1, we calculate $\eta$ and $\Gamma_\text{tail}$ both for the mesoscopic fluctuations in 3D, and Coulomb induced mesoscopic fluctuations in quasi-2D \cite{Skvortsov2013}. For $\eta$ the 3D contribution is given by \cite{Feigelman2012},
\begin{equation}
	\eta_\text{3D} = \frac{2\Delta_0}{\pi^2 \hbar D}\left(\frac{\rho_\text{N}}{R_\text{Q}}\right)^2 \ln\left(\frac{\xi}{l}\right),
\end{equation}
where $R_\text{Q} = 2\pi\hbar / e^2 = 25.8~\text{k}\Omega$ is the resistance quantum. For the $\beta$-Ta film, this results in $\eta_\text{3D} = 1.5\times10^{-5}$ and a gap broadening of $\Gamma_\text{gap}^\text{3D}/\Delta_0 = 1 - (1 - \eta_\text{3D}^{2/3})^{3/2}=8.9\times10^{-4}$.\\
The subgap tail in the density of states is due to the local variations of $\Delta$ \cite{Larkin1972,Feigelman2012}. In quasi-2D, homogeneously disordered films, it is always dominated by Coulomb-induced fluctuations \cite{Skvortsov2013,Feigelman2012}, which is given by,
\begin{equation}
	\frac{\Gamma_\text{tail}^\text{2D}}{\Delta_0} = \left(\frac{0.47}{g(g-g_\text{c})}\right)^{2/3},
\end{equation}
where $g=R_\text{Q} d / \rho_\text{N}$ and $g_\text{c} = \ln^2\left(\hbar/(k_\text{B}T_\text{c0}\tau_*)\right)/(2\pi)$, with $\tau_*=\max\{\tau_\text{e}, \tau_\text{e}(d/l)^2\}$. Here, $T_\text{c0}$ is the critical temperature without the presence of disorder. For the $\beta$-Ta film, we take $T_\text{c0}\approx T_\text{c}$. This is a reasonable approximation, since in that case $g\approx500$ is much larger than $g_\text{c}\approx 0.0058$ and $T_\text{c}$ only degrades significantly near the fermionic quantum critical point \cite{Finkelstein1987}. We obtain $\Gamma_\text{tail}^\text{2D}/\Delta_0 = 1.5\times 10^{-4}$.\\
The mesoscopic fluctuations in 3D without Coulomb interaction are given by \cite{Larkin1972,Skvortsov2013},
\begin{equation}
	\frac{\Gamma_\text{tail}^\text{3D}}{\Delta_0} = \left(\frac{5}{6^{1/4} (8\pi N_0 \Delta_0 \xi^3)^2}\right)^{5/4} \left(1 - \frac{\Gamma_\text{gap}^\text{3D}}{\Delta_0}\right),
\end{equation}
which results in $1.0\times10^{-6}$ for the $\beta$-Ta film. This is much smaller that the quasi-2D Coulomb contribution, and we therefore take $\Gamma_\text{tail}=\Gamma_\text{tail}^\text{2D}$.\\
With $\Gamma_\text{tail}^\text{2D}$, we can calculate the quasi-2D, Coulomb contribution to the gap broadening, which is given by \cite{Feigelman2012},
\begin{equation}
	\eta_\text{2D} = \left(1 - \left(1 - \frac{\Gamma_\text{tail}^\text{2D}}{\Delta_0}\ln^{2/3}\left(\frac{\Delta_0}{\Gamma_\text{tail}^\text{2D}}\right)\right)^{2/3}\right)^{3/2}.
\end{equation}
This results in $\eta_\text{2D}=9.0\times10^{-6}$, which is comparable to $\eta_\text{3D}$. We therefore take, $\eta = \eta_\text{2D} + \eta_\text{3D}$ \cite{Bespalov2016a}, which results in $\Gamma_\text{gap}/\Delta_0=1.2\times10^{-3}$.\\
The coherence peak energy $\Delta$ differs less than 0.1\% from $\Delta_0$ with this amount of disorder, which we calculated by numerically solving the Usadel equations with $\eta$ as pair-breaking parameter \cite{Driessen2012,Coumou2015}. That justifies the use of $\Delta_0$ in these calculations.

The density of states is thus altered by disorder which is parameterized by $\eta$ and $\Gamma_\text{tail}$, which we estimated above. Quasiparticles generated at $\Delta_0$ are mobile and relax to lower energies $\Delta-\Gamma_\text{gap}-\Gamma_\text{tail}$, where they localize \cite{Bespalov2016a}. At this energy, the length scale of the gap variations is $L_\text{T}=0.9\xi\left(\Gamma_\text{tail}/\Delta_0\right)^{-1/4}\approx 9\xi$ and within a volume $L_\text{T}^3$ the number of overlapping localized states is $N_\text{T}\simeq 0.46  N_0 \Delta_0 \xi^3 \left({\Gamma_\text{tail}}/{\Delta_0}\right)^{3/4}({\Delta_0}/{\Gamma_\text{gap}})$. Therefore, the quasiparticles will relax further until there is approximately one quasiparticle state per localization volume. The average distance between quasiparticles at this point is given by \cite{Bespalov2016a},
\begin{equation}\label{eq:rc}
	2 r_\text{c} \simeq 0.92 \xi \left(\frac{\Gamma_\text{tail}}{\Delta_0}\right)^{-1/4} \left(\ln N_\text{T}\right)^{-1/5}.
\end{equation}
This results in $r_\text{c}\simeq3.5\xi=65~\text{nm}$ for the $\beta$-Ta film and we estimate the number of localized quasiparticles states as $\tilde{n}_\text{qp}^\text{loc}=3/(4\pi r_\text{c}^3)\approx8.5\times10^{2}~\mu\text{m}^{-3}$. How many of these states are filled, depends on the (possibly non-equilibrium) agent that supports the localized quasiparticle density \cite{Bespalov2016a}. In our measurement, the dominating pair-breaking mechanism is thermal phonon absorption, see Supplementary Section 5.

\section{Estimation of read-out power effects on quasiparticle density and distribution function}
In all the calculations above, we assumed that the quasiparticles are in equilibrium with the thermal bath. This is supported by the observation that the measured variance in \cref{fig:results}(b) corresponds to thermal quasiparticle density fluctuations (the black line in \cref{fig:results}(b)). The continuous excitation with read-out power can however cause a non-equilibrium quasiparticle density at low bath temperatures \cite{deVisser2012a,Goldie2012,Fischer2023} and a non-thermal quasiparticle distribution over energy at higher bath temperature \cite{deVisser2014b,Goldie2012,Fischer2023}. \Cref{fig:Pread} shows the read-out power dependence of the relaxation time and variance. This is qualitatively consistent with the creation of excess quasiparticles: the relaxation time decreases and variance increases with increasing read-out power.\\
To limit these effects, we select the lowest read-out power at which we still have signal-to-noise ratio high enough to extract the relaxation time and variance (see Supplementary Section 4). This read-out power is the red data set in \cref{fig:Pread}, which is the same as the membrane data in \cref{fig:results}. We can estimate the bath temperature below which excess quasiparticles are expected from Eq. (35) of Ref. \cite{Fischer2023},
\begin{equation}
	T_\text{B}^* = \frac{(k_\text{B}T_*)^3}{k_\text{B}\Delta_0^2},
\end{equation}
where $k_\text{B}T_*$ is a measure for the width of the distribution function and is given by \cite{Fischer2023},
\begin{equation}
	\frac{k_\text{B}T_*}{\Delta_0} = \left(\frac{105\pi}{64}\left(\frac{k_\text{B}T_\text{c}}{\Delta_0}\right)^3  \frac{\hbar\tau_0}{\Delta_0^2} c_\text{phot}^\text{qp} P_\text{int}\right)^{1/6}.
\end{equation}
Here, $c_\text{phot}^\text{qp}=\alpha_\text{k}\omega_0/(2\pi N_0 V \Delta_0)$ is the photon-quasiparticle coupling constant and $P_\text{int} = 2 Q^2 P_\text{read} / (\pi Q_\text{c})$ is the internal microwave power in the resonator, with $Q\approx9.7\times10^3$ the loaded quality factor and $Q_\text{c}\approx 10.0\times10^3$ the coupling quality factor. With an on-chip read-out power of $P_\text{read}=-98~\text{dBm}$, this results in $T_\text{B}^*=19~\text{mK}$. For the substrate resonator (which we measure at $P_\text{read}=-100~\text{dBm}$), the same calculation results in $T_\text{B}^*=16~\text{mK}$. Since we measure at bath temperatures of $T>100~\text{mK}$, we do not expect excess quasiparticles generated by read-out power for the main data set (\cref{fig:results}). \\
The distribution function of the quasiparticles has the equilibrium Fermi-Dirac shape for energies larger than $E_* = \Delta_0 + k_\text{B}T_*\sqrt{T_*/T}$ \cite{Fischer2023}. In our measurement regime, we obtain $E_*= 1.3 \Delta_0$ and $1.4 \Delta_0$ for the highest and lowest bath temperatures. For lower energies than $E_*$, the quasiparticle distribution function could deviate from the thermal distribution, and the average quasiparticle energy could be slightly higher than $\Delta_0$ (when a BCS density of states is assumed, as is done in these calculations). At the same time, we calculated the recombination and scattering times for a quasiparticle at $\Delta_0$ in \cref{eq:trec_e-ph,eq:tscat_e-ph2D,eq:tscat_e-ph3D}. A higher quasiparticle energy of $1.4\Delta_0$ (which is a strict upper limit to the average quasiparticle energy) will however only result in a faster recombination time by less than a factor 2 \cite{Kaplan1976} and the exponential temperature dependence ($\tau_\text{rec}\propto 1/n_\text{qp}$) is preserved in the case of a non-equilibrium distribution\cite{Goldie2012}.\\
Moreover, the enhanced electron-electron inelastic relaxation due to disorder (see \cref{fig:lifetimes}) will result in a more thermal shape of the distribution function than calculated by Refs. \cite{Goldie2012,Fischer2023}, which only consider electron-phonon interactions. Additionally, the broadened density of states due to disorder (see Supplementary Section 6) will also result in a broadening of the non-equilibrium peaks in the distribution function and thus diminish the read-out power effects on the distribution function.

Therefore, at these low read-out powers, we expect a thermal quasiparticle density and an approximately thermal shape for the distribution function, with an average quasiparticle energy close to the gap energy. This justifies the relaxation time calculations in Supplementary Section 1. 

\clearpage
\bibliography{bTa_lifetimes}
\end{document}